\documentclass[%
 preprint,
 superscriptaddress,
 amsmath,amssymb,
 aps,physrev,
]{revtex4-2}

\usepackage{graphicx}
\usepackage{dcolumn}
\usepackage{bm}
\usepackage{xcolor}
\usepackage{amsmath}
\usepackage{placeins}   
\usepackage{float}
\def\unit#1{\ensuremath{\mathrm{\,#1}}}
\def\celsius{\ensuremath{\,^{\circ}\mathrm{C}}}
\newcommand{\MM}{Sec.~Materials \& Methods}
\newcommand{\SuppInfo}{Supporting Information}

\begin{document}

\title{Simultaneous measurements of enzyme propagation and biopolymer mobility elucidate the directional degradation of a dense biopolymer matrix}

\author{Vincenzo Ruzzi}
\altaffiliation{Present address: Faculty of Physics, University of Vienna, Boltzmanngasse 5, 1090 Vienna, Austria}
\affiliation{Laboratoire Charles Coulomb (L2C), CNRS, Universit\'e de Montpellier, Montpellier, France}

\author{Antoine Bouchoux}
\affiliation{Toulouse Biotechnology Institute (TBI), Universit\'e de Toulouse, CNRS, INRAE, INSA, Toulouse, France}

\author{Carole Antoine-Assor}
\affiliation{IATE, INRAE, Institut SupAgro, Universit\'e de Montpellier, Montpellier, France}

\author{Donna-Joe Bigot}
\affiliation{Toulouse Biotechnology Institute (TBI), Universit\'e de Toulouse, CNRS, INRAE, INSA, Toulouse, France}

\author{Salma Menzeh}
\affiliation{Toulouse Biotechnology Institute (TBI), Universit\'e de Toulouse, CNRS, INRAE, INSA, Toulouse, France}

\author{Maike Petermann}
\affiliation{Toulouse Biotechnology Institute (TBI), Universit\'e de Toulouse, CNRS, INRAE, INSA, Toulouse, France}

\author{Laurent Leclercq}
\affiliation{IBMM, Universit\'e de Montpellier, CNRS, ENSCM, Montpellier, France}

\author{Herv\'e Cottet}
\affiliation{IBMM, Universit\'e de Montpellier, CNRS, ENSCM, Montpellier, France}

\author{C\'edric Montanier}
\affiliation{Toulouse Biotechnology Institute (TBI), Universit\'e de Toulouse, CNRS, INRAE, INSA, Toulouse, France}

\author{Claire Dumon}
\affiliation{Toulouse Biotechnology Institute (TBI), Universit\'e de Toulouse, CNRS, INRAE, INSA, Toulouse, France}

\author{Luca Cipelletti}
\affiliation{Laboratoire Charles Coulomb (L2C), CNRS, Universit\'e de Montpellier, Montpellier, France}
\affiliation{Institut Universitaire de France, Paris, France}

\author{Laurence Ramos}
\email{laurence.ramos@umontpellier.fr}
\affiliation{Laboratoire Charles Coulomb (L2C), CNRS, Universit\'e de Montpellier, Montpellier, France}

\date{\today}



\begin{abstract}
Enzymatic degradation of dense biopolymer matrices is central to natural processes like digestion and cancer progression, as well as applications in biomedicine, biomass valorization, and nanotechnology. Yet, it remains unclear how enzyme propagation and substrate degradation interact, and especially whether and how enzymatic activity shapes the propagation process. Here, we bridge this gap by employing Fluorescence-Photon Correlation Imaging to simultaneously track directional enzyme propagation and biopolymer degradation in a model system. We use viscoelastic solutions of arabinoxylan, a hemicellulosic polysaccharide with a xylan backbone substituted by arabinose and ferulic acid, and a fluorescently labeled arabinoxylanase, an enzyme that randomly cleaves the xylan chain. To resolve enzyme propagation and biopolymer degradation with spatiotemporal precision, we combine fluorescence imaging of the enzyme with speckle-based photon correlation imaging, a technique to quantify biopolymer mobility, which we show is a proxy for biopolymer degradation. Our experiments reveal that enzyme diffusion is decoupled from catalytic activity, with no evidence of activity-enhanced diffusion. Instead, the progression of the degradation front is governed by a combination of enzyme diffusion, reaction kinetics, and slow enzyme deactivation, the latter being characterized by a double-exponential decay. By extending the Michaelis-Menten framework to space- and time-dependent conditions, we quantify how cumulative enzyme exposure and deactivation shape the degradation dynamics. This work establishes a quantitative framework to understand enzymatic processes in concentrated, beyond-dilute-regime substrates, with implications for the rationalization of biocatalytic systems in industrial and biomedical applications.
\end{abstract}

\maketitle
\newpage
\section*{Introduction}
Enzymatic degradation of biopolymers is a natural phenomenon of macromolecular deconstruction occurring in several biological processes, such as food digestion~\cite{chi2022}, cell lysis~\cite{salazar2007}, and cancer progression~\cite{garcia2017, cox2011, oskarsson2013, garcia2017}. More recently, enzymatic degradation has been engineered in a wide range of applications. These include the valorization of plant biomass for the sustainable production of biofuels and bioplastics from organic waste~\cite{pauly2010, heux2015, brodin2017, hassan2019, ning2021}, as well as controlled drug delivery, where the targeted decomposition of a carrier can be triggered by injectable enzymes~\cite{minehan2022, andresen2010, gameel2025, ma_bespoke_2025}. In most of these cases, degradation is directional, progressing as a moving front driven by the enzyme concentration gradient that develops as enzymes diffuse into the biopolymer matrix, e.g., as in the case of wheat straw~\cite{blosse2023}. Thus, not only does the efficiency of enzymatic action depend on enzyme activity, it also depends on the ability of enzymes to penetrate and advance through the substrate, which typically consists of a viscoelastic matrix, such as a concentrated polymer solution or a gel.  Whether and how enzymatic activity influences enzyme propagation remains however unclear. Establishing a quantitative link between these two aspects requires monitoring simultaneously \textit{in-situ} the degradation of the substrate and the propagation of the enzymes, which challenges the currently available approaches.

Enzyme diffusivity is commonly measured using fluorescently-labeled enzymes in simple dilute solutions with fluorescence-based techniques such as Fluorescence Recovery After Photobleaching (FRAP) and Fluorescence Correlation Spectroscopy~\cite{muddana2010, jee2018, jee2020, yancheva2025, fadda2003, fong2016}. Notably, some of these studies suggest that the diffusivity of degrading enzymes in simple molecular solutions is higher than that in the absence of the substrate~\cite{muddana2010, jee2018, jee2020}. This boost effect, also termed enhanced enzymatic diffusion (EED), is still highly debated~\cite{feng2020, zhang2021, jee2024, yancheva2025}. However, investigations of enzyme mobility in semi-dilute or concentrated biopolymer matrices remain relatively limited. Most available studies in crowded systems rely on FRAP measurements and use fluorescent dextran polymers or inert proteins, such as bovine serum albumin, as surrogates for enzymes, because they have comparable size~\cite{paes2012, paes2013}. This approach is largely motivated by the difficulty of labeling enzymes with fluorophores while preserving their catalytic activity, a challenge in studies addressing EED~\cite{jee2018}. Although one FRAP-based study~\cite{cuyvers2011} reported a higher diffusivity for an active enzyme interacting with a polymeric substrate compared with its inactive counterpart, the underlying mechanisms could not be conclusively identified because concomitant measurements to quantify the degradation of the biopolymer matrix were lacking. 

A complementary approach aims at characterizing structural and physico-chemical changes occurring within the biopolymer matrix as degradation progresses~\cite{weetall1993, cheng2000, lairez2007, pasquier2019, bayrak2021, napieraj2024, blosse2023}. However, in these studies, primarily relying on synchrotron radiation, light scattering, or rheology, data on enzyme motion are entirely absent. Consequently, the relationship between enzyme transport and the resulting modifications of the polymer matrix remains poorly understood. This limitation becomes particularly critical in directional degradation processes, where enzyme transport and substrate transformation are intrinsically coupled in space and time. This aspect is not addressed in the works cited above, where enzymes were always homogeneously distributed in the bulk matrix. Degradation directionality, and the related question of enzyme propagation kinetics in degradable biopolymer systems, has only begun to be addressed in the past decade with experimental studies using small-angle X-ray scattering~\cite{napieraj2024} or combining deep-UV fluorescence microscopy and imaging~\cite{tawil2011, jamme2014, devaux2018, bonnin2019, voisin2025}. However, these works probe either the propagation of the enzymes in mm-sized samples by consecutive imaging scans with poor temporal resolution~\cite{bonnin2019, voisin2025} or the results of their action (e.g. change in polymer structure)~\cite{napieraj2024}, whereas a full understanding of the process requires simultaneous, spatially- and temporally-resolved measurements of both enzyme propagation and biopolymer degradation. 

Here, we address these challenges by measuring simultaneously the directional enzymatic degradation of a biopolymer and the propagation of the enzymes through the biopolymer matrix. We use a model biopolymer/enzyme pair comprising arabinoxylan (AX), a water-soluble hemicellulosic polysaccharide with a linear backbone, here used in its feruloylated form (FAX~\cite{carvajal2005, antoine2021}), and an endo-xylanase that randomly cleaves the xylan backbone~\cite{vardakou2008} (details in \MM). The water solubility of FAX, its linear backbone, and ability to form concentrated polymer solutions make it an ideal model for investigating enzyme transport and degradation in dense polymer matrices. FAX is also relevant to applications such as biomass valorization~\cite{heux2015, bilal2024} and drug delivery~\cite{perez2024}. Photon Correlation Imaging~\cite{duri2009}, a method that blends imaging and dynamic light scattering, is used to measure with spatial and temporal resolution the microscopic biopolymer dynamics, which is directly related to the biopolymer degradation. Concomitantly, we quantify the time-evolving concentration profile of the fluorescently-tagged enzyme by fluorescence imaging. We show that enzyme diffusion is insensitive to catalytic activity, ruling out any boost effect in our experiments. By extending the Michaelis-Menten equation for substrate degradation~\cite{marangoni2003} to space-varying conditions, we rationalize the advancement of the degradation front, which is governed by enzyme diffusion, reaction kinetics, and time-dependent enzyme deactivation. This work establishes a quantitative framework to understand and optimize enzymatic degradation in conditions relevant for applications~\cite{akram_insight_2025,hlaing_enzymatic_2026,huang_unleashing_2026}.


\section*{Experimental Results}\label{sec:res}

\begin{figure}[t!]
    \centering
    \includegraphics[width=\columnwidth]{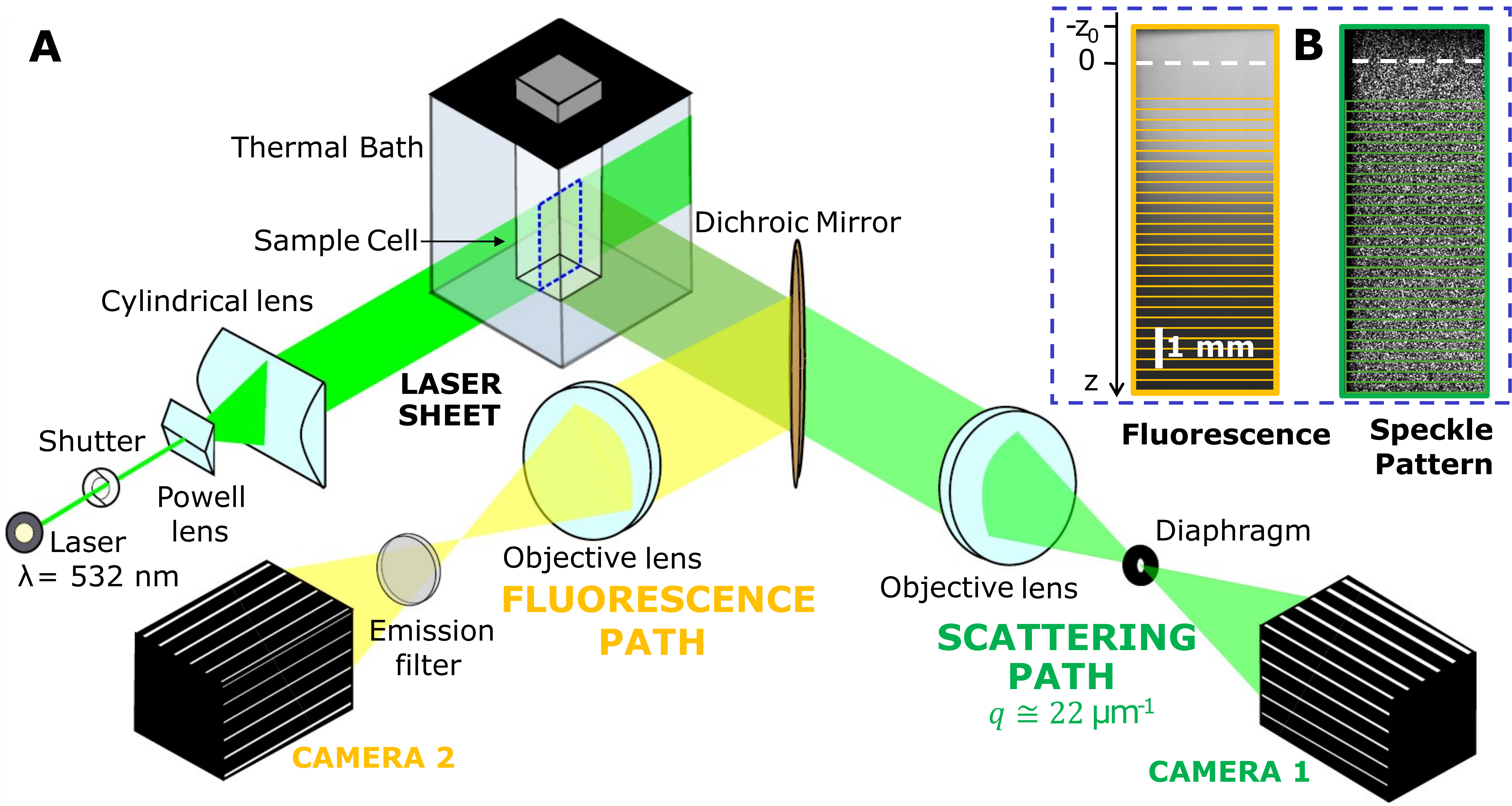}
    \caption{\textbf{Experimental setup.} (A)~Scheme of the Fluorescence - Photon Correlation Imaging setup. The sample volume illuminated by the incident beam is highlighted by the dashed blue box and is imaged, with magnification one, on cameras 1 and 2. The scattering and the fluorescence paths are depicted in green and yellow, respectively. (B)~Typical fluorescence (left) and scattering (right) images, divided into $28$ Regions of Interest (ROIs), for calculating the $z$-dependent enzyme concentration and polymer mobility, respectively. The contrast of the scattering image has been enhanced to better visualize the speckle pattern. The scale bar is the same for the two images. The white dashed horizontal lines at $z=0$ indicate the position of the interface between the enzyme solution and the polymer solution at time $t=0$; $z= -z_0$ is the air-sample interface after depositing the enzyme solution on top of the polymer solution.}
    \label{fig:1}
\end{figure}
\subsection*{Experimental protocol} We briefly describe the system and setup, see \MM~for details. The substrate, Feruloylated Arabinoxylan (FAX), is a linear neutral polysaccharide with arabinosyl-feruloylated side groups. It is diluted in good solvent conditions in an aqueous buffer at pH = 5~\cite{carvajal2005}. The FAX concentration, $c$, is either 10 or 20 g/L, significantly larger than the overlap concentration $c^* \simeq 2 \unit{g/L}$, hence corresponding to the semi-dilute visco-elastic regime~\cite{carvajal2005}. The enzyme is a xylanase, which randomly cleaves the polysaccharide backbone~\cite{vardakou2008}, reducing its molecular mass, which results in enhanced microscopic mobility of the FAX chains. In a typical experiment, a sample cell with a rectangular cross section of $4\times 10 \unit{mm^2}$ is initially filled with $420 \unit{\mu L}$ of FAX solution up to a height of 10.5 mm. At time $t = 0$, $70 \unit{\mu L}$ of enzyme solution are gently poured on top of the  FAX solution, after which we simultaneously probe the evolution of the biopolymer degradation and enzyme propagation along the vertical $z$-direction. We vary the enzyme activity and its concentration,  $0.55\unit{\mu M}  \le [E]_{\infty} \le 2.2 \unit{\mu M}$, with $[E]_{\infty}$ the enzyme concentration averaged over the total sample volume. We mainly use a wild type (WT) enzyme, but some experiments are also conducted with a mutant enzyme, whose activity is significantly reduced as compared to that of the WT enzyme (see Sec.~Materials \& Methods and~\SuppInfo~for details). The investigated parameter space is detailed in Table~\ref{table:samples} in Section Materials \& Methods. 

We use a Fluorescence-Photon Correlation Imaging (Fluo-PCI) apparatus (Fig.~\ref{fig:1}), specifically designed to probe simultaneously with space- and time-resolution the biopolymer degradation and the enzyme propagation. The apparatus consists of a PCI setup~\cite{duri2009, secchi2013, usuelli2022, milani2024} coupled to a fluorescence imaging line. A laser sheet (wavelength $\lambda_0 = 532.5 \unit{nm}$) impinges on the sample cuvette, which is immersed in a transparent water bath to control temperature. To separate out the contributions of the biopolymer solution and fluorescent enzyme, the light scattered at $\theta = 90$ deg and the fluorescence emission are split by a dichroic mirror into two detection paths that form an image of the sample volume. A speckled image of the scattering volume is collected by camera 1, to measure the microscopic dynamics of the polymer. The intensity of the fluorescence light, proportional to the enzyme concentration, is collected by camera 2.

\subsection*{Degradation front investigated by Photon Correlation Imaging}

\begin{figure}[t!]
    \centering
    \includegraphics[width=\columnwidth]{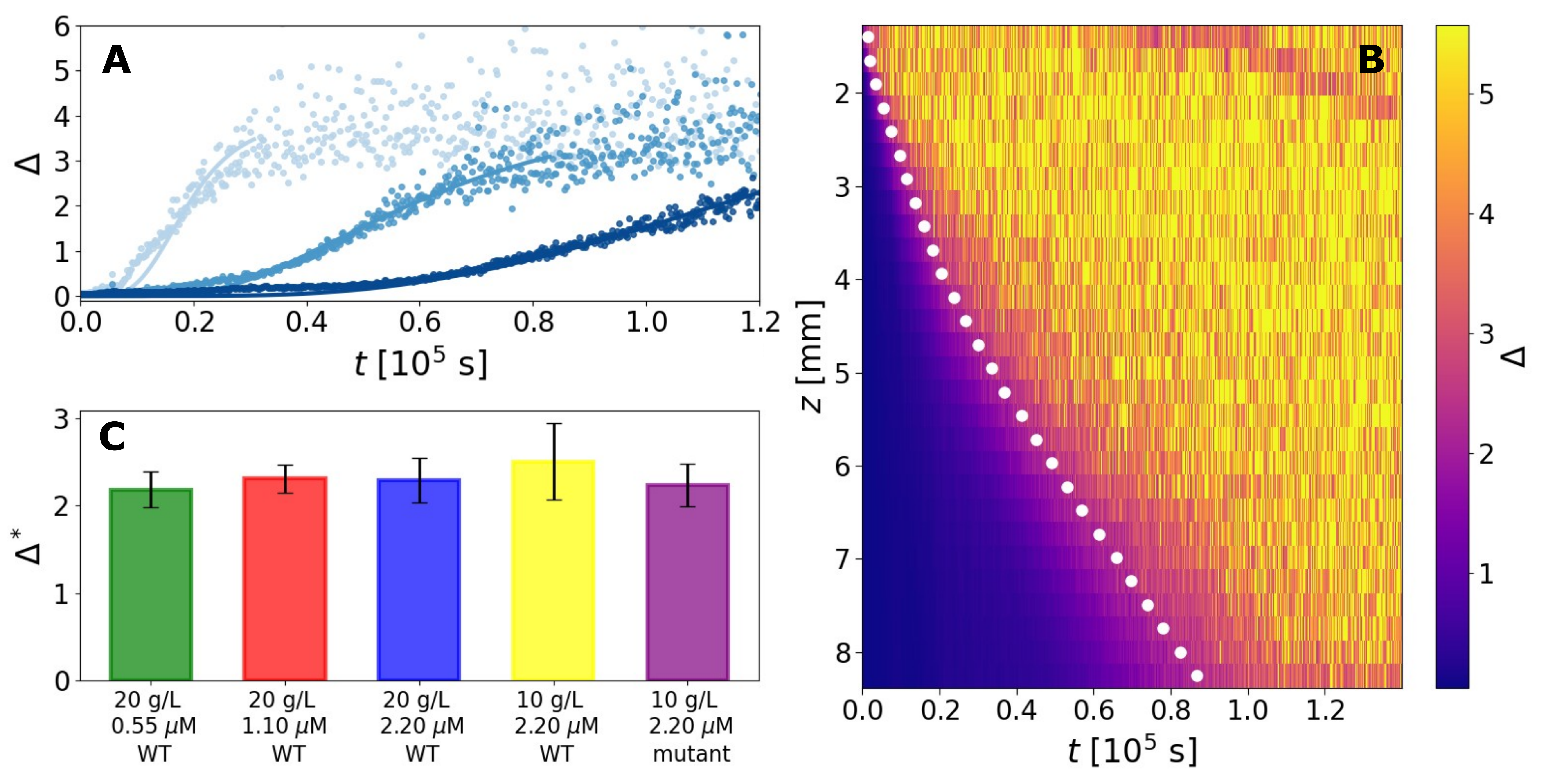}
    \caption{\textbf{Spatio-temporal evolution of the microscopic mobility $\Delta$ of a biopolymer solution undergoing enzymatic degradation}. (A)~Time evolution of $\Delta$ at three depths in the biopolymer matrix, $z_1 = 2.92 \unit{mm}$, $z_2 = 5.46 \unit{mm}$, $z_3 = 8.00 \unit{mm}$, from light to dark blue. ($c = 20 \unit{g/L}$, $[E]_{\infty} = 2.2 \unit{\mu M}$ WT). The full lines are best fits using the model discussed in the Modeling section below, see Eq.~\ref{eq:MM_for_mobility_modified}. (B)~Mobility kymograph $\Delta (z,t)$ throughout degradation: Violet zones correspond to a low-mobility non-degraded solution, yellow regions indicate a high-mobility degraded state. The white dots superimposed to the map show the degradation front points $(z^*, t^*)$, as defined in the text. (C)~Microscopic mobility $(\Delta^* \pm \delta \Delta^*)$ at the degradation front for all investigated samples. The average over all conditions is $\langle \Delta^* \rangle = 2.31 \pm 0.29$.}
    \label{fig:2}
\end{figure}

We first focus on the spatio-temporal evolution of the microscopic dynamics of the biopolymer solution during degradation, which is quantified by a dimensionless polymer mobility $\Delta$, obtained by PCI (see Materials \& Methods), and directly related to the concentration of the product P resulting from the enzymatic reaction, as we shall show it. $\Delta = 1$ corresponds approximately to polymer chain displacements of $\sim 80 \unit{nm}$ over a time interval $\Bar{\tau} = 20 \unit{ms}$.
Figure~\ref{fig:2}A shows the time evolution of $\Delta$ for three representative Regions of Interest (ROIs) at depths $z_1 = 2.92 \unit{mm}$, $z_2 = 5.46 \unit{mm}$, and $z_3 = 8.00 \unit{mm}$, from light blue to dark blue, for a sample with $c = 20 \unit{g/L}$ and WT enzyme at $[E]_{\infty} = 2.2 \unit{\mu M}$. The mobility increases from low values ($\Delta \ll 1$) for the pristine biopolymer up to $\Delta \simeq 4$ in the degraded state, corresponding to displacements larger than about $200 \unit{nm}$, the upper detection limit of our apparatus (see~\SuppInfo). As the degradation starts from the upper layers of the biopolymer matrix, small $z$, the deeper in the sample, i.e., the larger $z$, the longer the characteristic time for the increase of $\Delta$. 

The whole degradation process may be visualized by constructing a mobility kymograph, Fig.~\ref{fig:2}B, in which the evolution of $\Delta(z,t)$ is displayed as a function of experimental time (horizontal axis), and depth in the sample (vertical axis). The kymograph reveals low-mobility spatio-temporal areas (purple), corresponding to the non-degraded biopolymer, and high-mobility areas (yellow), corresponding to degraded biopolymer. Accordingly, the histogram of the mobility measured over the full kymograph exhibits two distinct peaks. This allows for setting unambiguously a mobility threshold  $\Delta^*$, using the Otsu algorithm~\cite{otsu1979} (see Materials and Methods and Fig.~S4 in the~\SuppInfo   ~for details). 
The set of points $(z^*,t^*)$ such that $\Delta(z^*,t^*) = \Delta^*$ identifies the spatio-temporal degradation front, shown by the white dots in Fig.~\ref{fig:2}B. We repeat the same procedure for various experimental conditions (biopolymer and enzyme concentrations, active WT enzyme or weakly active mutant enzyme), finding similar $\Delta^*$ values (Fig.~\ref{fig:2}C), supporting the robustness of this procedure and suggesting similar degradation conditions at the front.

\begin{figure}[]
    \centering
    \includegraphics[width=0.5\columnwidth]{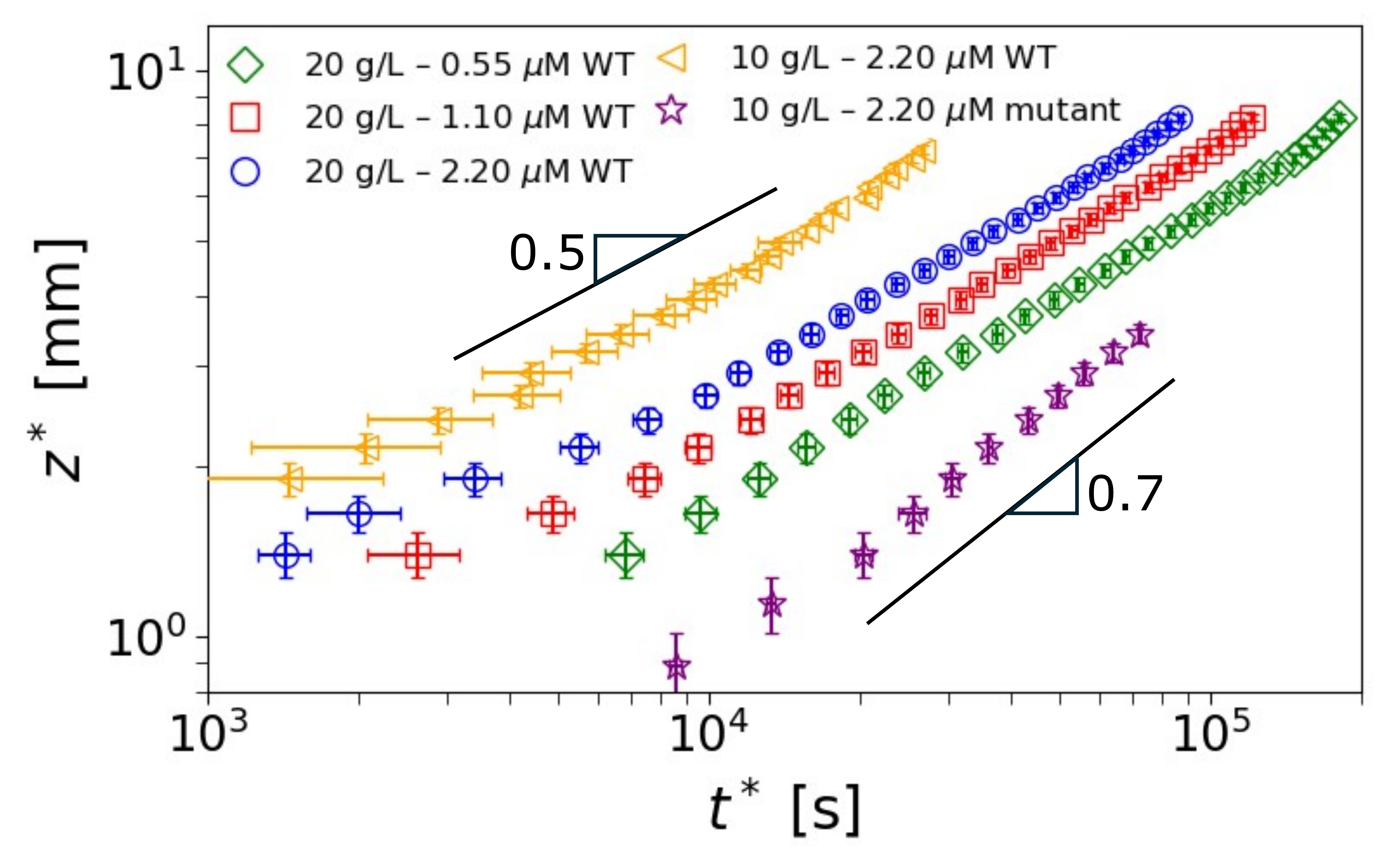}
    \caption{\textbf{Propagation of the degradation front}. Double-logarithmic plot of the position of the degradation front as a function of time, for all experimental conditions. The front propagation is close to a power law, $z^* \sim t^b$, with $b \simeq 0.5$ for the WT enzyme and $b \simeq 0.7$ for the mutant enzyme.}
    \label{fig:3}
\end{figure}

The front propagation, by contrast, strongly varies with experimental conditions. Figure~\ref{fig:3} shows in a double logarithmic scale the front position $z^*$ as a function of $t^*$, the degradation front time. For all experiments performed with the WT enzyme, $z^*$ varies approximately as a power law with time: $z^* = a(t^*)^b$, with a same exponent $b = 0.515 \pm 0.003$, but with very different prefactor $a$. We find that, for a same enzyme concentration, the front propagates faster when the biopolymer concentration is lower, compare the yellow symbols for $c=10$ g/L and the blue symbols for  $c=20$ g/L in Fig.~\ref{fig:3}. Furthermore, for a given biopolymer concentration ($c=20$  g/L), the degradation front is slower for lower enzyme concentrations: $a$ decreases by a factor of $1.6$, as $[E]_{\infty}$ decreases from $2.2 \unit{\mu M}$ to $0.55 \unit{\mu M}$.  Finally, the advancement of the degradation front is significantly slower for the mutated enzyme than for the wild type one, compare the purple stars and the yellow triangles in Fig.~\ref{fig:3}, both referring to $c=10$ g/L and $[E]_{\infty}=2.2 \unit{\mu M}$. Thus, our experimental observations indicate that the kinetics of the degradation process is governed not only by the intrinsic enzyme activity, but also by the ratio between the enzyme and polymer concentrations. 

Interestingly, the exponent $b \simeq 0.5$ measured for the WT enzyme is reminiscent of the scaling law for a process ruled by diffusion. However, we will show below that this is purely fortuitous: in general, the front propagation depends both on enzyme propagation through the matrix and on the degradation reaction, as illustrated by the different scaling exponent found for the mutated enzyme, $b \simeq 0.7$. To reach a complete mechanistic understanding of the processes at play and unveil the relevant biochemical and physical mechanisms, it is therefore crucial to independently probe how the enzymes propagate in the biopolymer matrix.

\subsection*{Enzyme propagation investigated by fluorescence imaging}

\begin{figure}
    \centering
    \includegraphics[width=0.5\columnwidth]{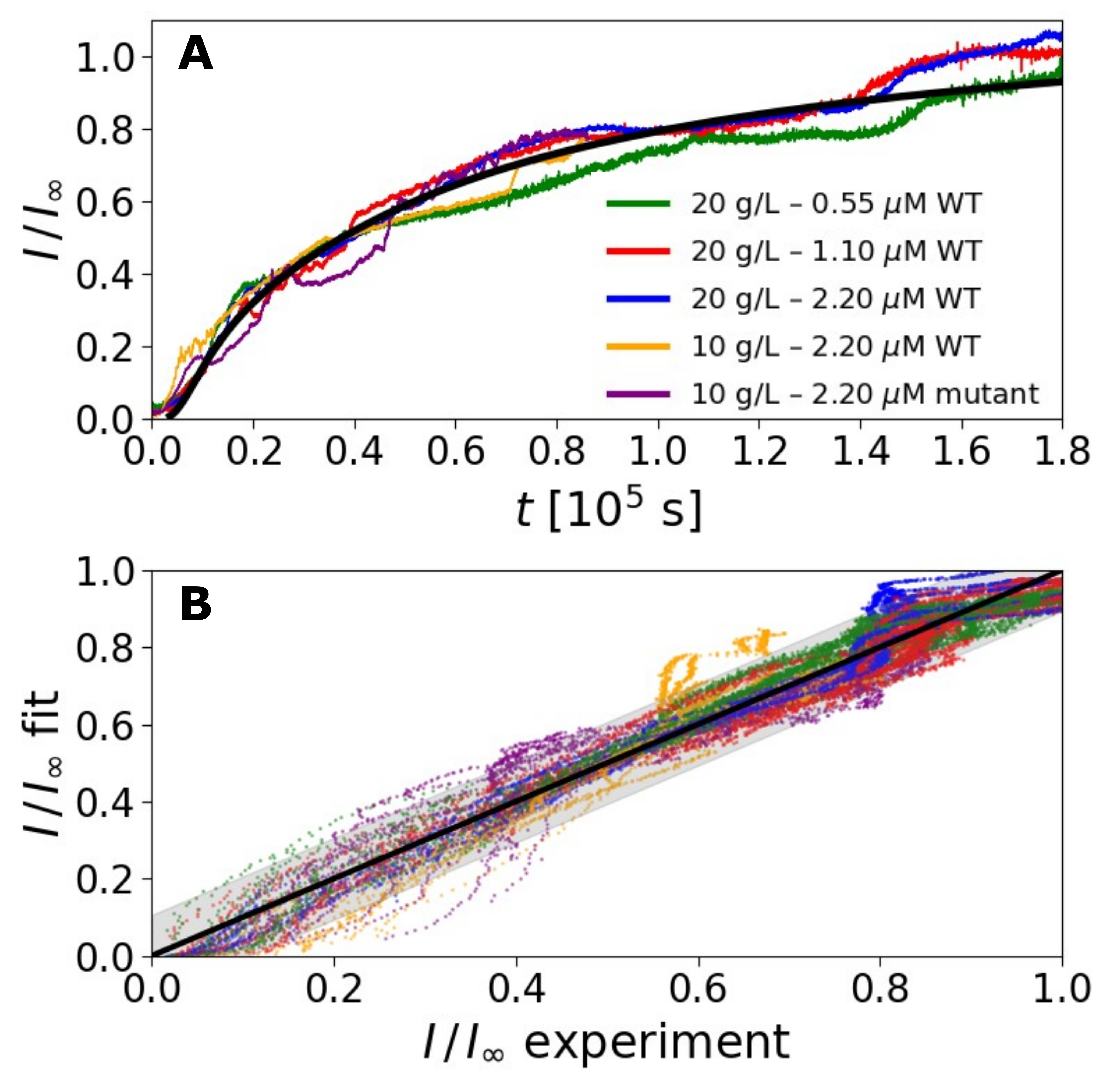}
    \caption{\textbf{Enzyme propagation}. (A)~Time evolution of the normalized fluorescence intensity $\Tilde{I} = I/I_{\infty}$ for samples with various biopolymer concentrations, enzyme concentrations and types of enzymes, at $z= 6.22 \unit{mm}$. The black line is a global fit of the data accounting for the diffusion of the labeled enzyme in equilibrium with a fraction of free fluorophore, see text for details. (B)~Fitted $\tilde{I}$ \textit{vs} the corresponding experimental data, for all the samples of panel A (same color code), and for depths $z = 2.92, \, 3.68, \, 4.45, \, 4.95, \, 5.46, \, 6.22, \, 6.99, \,$ and $7.49 \unit{mm}$. The black straight line corresponds to the expected behavior according to the model and the gray-band indicates the standard deviation of the residuals of the global fit, $\sigma_I = 0.1$.}
    \label{fig:4}
\end{figure}

The temporal evolution of the enzyme concentration profiles is obtained from the fluorescence signal measured by the Fluo-PCI setup. The signal stems from the fluorescently-labeled enzyme and, possibly, from free fluorophore molecules. Figure~\ref{fig:4}A shows, for all experimental conditions, the normalized fluorescence intensity, $\Tilde{I} = I/I_\infty$, \textit{vs} time, at depth $z= 6.22 \unit{mm}$, where $I_\infty$ is the intensity value asymptotically reached at long times, typically $t > 2 \times 10^5 \unit{s}$. Remarkably, the increase of $\Tilde{I}$ with time is the same regardless of polymer and enzyme concentrations and enzyme activity (WT \textit{vs} mutant enzymes). The fluorescent intensity evolution can be modeled assuming diffusive behavior, using the analytical solution~\cite{crank1979} for the one-dimensional problem corresponding to our experimental configuration, where the concentration profile only varies with depth $z$. Both the contribution of the labeled enzyme and that of free fluorophore molecules have to be considered: we find that the quality of a one-component modeling is indeed statistically poorer, see Materials \& Methods and the~\SuppInfo~for details.
Data for all samples and depths $2.92 \leq z \leq 7.49 \unit{mm}$ are very well reproduced by the model, Eqs.~\ref{eq:diffusion_solution}--\ref{Eq:Cranck1} of the Materials \& Methods, as shown by the black lines in Figs.~\ref{fig:4}A,B. The top-most ($z < 2.92  \unit{mm}$) layer is excluded from the fit, due to the unavoidable local mixing when injecting the enzyme solution. The lowest available $z$ slightly varies from run to run, depending on illumination: here we restrict the fitting range to the lowest $z$ shared by all runs. 

Since the intensity evolution is nearly independent of experimental conditions, the fit parameters are shared by all datasets. The fit parameters are $D_l = (1.02\pm 0.37) \times 10^{-10} \unit{m^2/s}$, $D_s = (6.59 \pm 0.77) \times 10^{-10} \unit{m^2/s}$, and $r = 0.47\pm 0.03$, where $D_l$, $D_s$ are the diffusion coefficients of the larger and smaller species, respectively, and $r$ is the intensity-weighted relative number of the larger diffusing species. Using the Stokes-Einstein relation, $R_{l,s}=k_B T/(6\pi\eta D_{l,s})$, we calculate the hydrodynamic radius $R_{l,s}$ of each species. Here, $k_B$ is the Boltzmann's constant, $T$ the absolute temperature, and $\eta$ the viscosity of the medium in which the enzyme and the fluorophore diffuse. Since $D_{l,s}$ is independent of polymer concentration and hence of the macroscopic viscosity of the solutions, we hypothesize that the enzyme and the fluorophore diffuse freely through the pores of the polymer solution, experiencing the viscosity of the solvent at the working temperature of $37^\circ$C, $\eta = 0.69$~mPa~s. With this assumption, we find $R_l \simeq 3.2 \unit{nm}$, in good agreement with the expected size of the enzyme~\cite{vardakou2008}, and $R_s \simeq 0.5 \unit{nm}$, corresponding to the size of the free fluorophore molecules. Independent measurements of the labeled enzyme solution by Taylor Dispersion Analysis confirm both the presence of free fluorophore in equilibrium with the labeled enzyme and the values of $R_{l,s}$ obtained from the fits shown in Fig.~\ref{fig:4}A (see~\SuppInfo ~for details).

Our findings demonstrate that the enzyme can freely diffuse through the biopolymer matrix, whose typical pore size, $\xi \simeq 30 \unit{nm}$ at $c = 20 \unit{g/L}$, is significantly larger than $R_l$, see Section Materials \& Methods. This is consistent with scaling theories for the diffusion of non-sticky tracers in polymer solutions~\cite{cai_mobility_2011}. Moreover, because the mutated enzyme shows the same behavior as the wild type one, we conclude that the diffusive propagation of the enzymes in the biopolymer matrix is not affected by their activity. Thus, in our experimental conditions, we can exclude enhanced enzymatic diffusion, a topic intensively debated in the community~\cite{feng2020, zhang2021, jee2024, yancheva2025}.

\section*{Modeling}\label{sec:disc}

Combining the spatio-temporal maps of both enzyme concentration and biopolymer mobility allows for testing various scenarios for the coupling between enzyme propagation and substrate degradation. Since the fluorescence measurements indicate that the enzyme diffuses freely through the matrix, independently of its activity, a simple picture would be that the degradation front corresponds to a given enzyme concentration. We test this hypothesis by measuring $[E]^* = [E](z^*,t^*)$, the enzyme concentration at the degradation front, for three experiments with fixed biopolymer concentration $c = 20 \unit{g/L}$, and different WT enzyme concentration $[E]_{\infty}$. We find that, for a given $[E]_{\infty}$,  $[E]^*$ is roughly constant across the set of pairs $(z^*,t^*)$ that define the degradation front, Fig.~S8,~\SuppInfo. 
However, $[E]^*$ varies significantly with $[E]_{\infty}$. Indeed,   
$\langle [E]^* \rangle$, the enzyme concentration averaged over the front position ($2.92~$mm~$\le z^* \le 8.25$~mm), decreases from  $(0.69 \pm 0.05) \unit{\mu M}$ to $(0.39 \pm 0.02) \unit{\mu M}$ as $[E]_{\infty}$ decreases from $2.2$ to $0.55 \unit{\mu M}$. This rules out the simple hypothesis formulated above and calls for a full modeling of the non-stationary, directional kinetics of degradation.

\begin{figure}
    \centering
    \includegraphics[width=0.5\columnwidth]{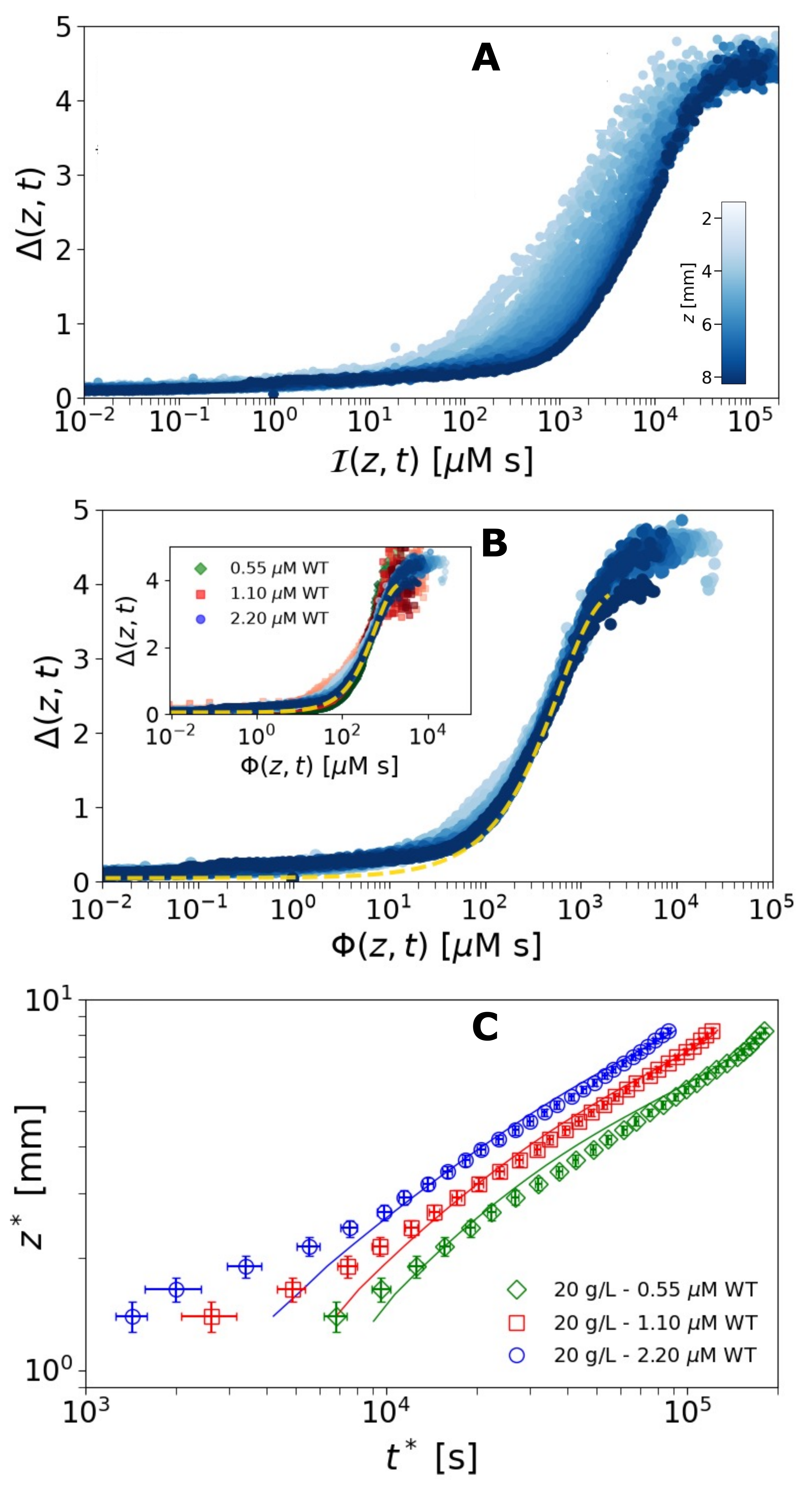}
    \caption{\textbf{Modeling of the degradation front considering diffusing enzymes with a time-dependent deactivation.} (A)~Mobility $\Delta$ as a function of the integral of the enzyme concentration at depth $z$ from time $0$ to $t$, $\mathcal{I}(z,t)$, for samples with biopolymer concentration $c = 20 \unit{g/L}$ and  wild type enzyme concentration $[E]_{\infty} = 2.2 \unit{\mu M}$. Each curve corresponds to a given $z$, from $z= 2.92$ to $z=8.00 \unit{mm}$ (see color bar). (B)~$\Delta$ as a function of $\Phi (z,t) = \int_0^t [E](z,t')f(t')\mathrm{d}t'$ with the same color-code as in (A), where the function $f(t')$ accounts for enzyme deactivation, see main text. Inset:~$\Delta$ as a function of $\Phi (z,t)$ for $[E]_{\infty} = 0.55, 1.1$ and $2.2 \unit{\mu M}$ (see legend). 
    Data collected at different $z$'s (main panel: same color code as in panel A) nearly collapse on the same curve, in agreement with the model (yellow dashed line, $z=5.21 \unit{mm}$). 
    (C)~Degradation fronts at $c = 20 \unit{g/L}$ from experiments (symbols) and from the calculated mobility maps (lines) for different enzyme concentrations.
    }
    \label{fig:5}
\end{figure}

We start from the classical Michaelis-Menten (MM) approach, which describes an enzyme-catalysed reaction involving the transformation of a substrate S into a product P~\cite{marangoni2003}. In our case, the substrate is the xylosidic (X-X) bond of the FAX chains and the products are mainly short-length FAX oligosaccharides~\cite{rudjito2023}. The backbone of the FAX biopolymer contains approximately $2700$ X-X bonds~\cite{antoine2021}, yielding for a FAX concentration $c = 20 \unit{g/L}$ an estimate of $80 \unit{mM}$ X–X bonds, which corresponds to the initial substrate concentration $[S]_0$. This is several orders of magnitude larger than the enzyme concentration: We thus assume that the quasi-steady-state approximation for the enzyme-substrate complex holds throughout our experiments, i.e., that the reaction is not limited by the exhaustion of the substrate. With this assumption, the kinetics of the enzymatic degradation is ruled  by a differential equation that describes the time-dependent decrease (respectively, increase) of the substrate concentration $[S]$ (respectively, product concentration $[P]$), see \MM, Eq.~\ref{MM_equation_classic}.

Our experiments differ from the conventional modeling of an enzymatic reaction via the MM formalism in three respects: i) we measure the substrate mobility $\Delta$, rather than the product concentration $[P]$; ii) the experiments probe degradation over extended periods of time; iii) they are run under spatially heterogeneous conditions. To address i), we show in separate tests that $\Delta$ is a proxy for $[P]$, since $[P] = A\Delta +B$, with $A$ and $B$ constants, see Fig.~S3 
of the~\SuppInfo. Hence, a MM-like equation, Eq.~\ref{MM_for_mobility}, holds also for $\Delta$, allowing for a direct comparison with our experimental data. Given ii), we model the full time evolution of [P] (or $\Delta$), rather than restricting the analysis to the initial reaction rates, as in conventional MM studies. To account for iii), we extend the conventional MM formalism by imposing that $\Delta$ and the enzyme concentration $[E]$ depend on both time and depth $z$, see Eq.~\ref{MM_for_mobility}. Note that in the model we have assumed that the concentration profile of P (or, equivalently, the mobility profile $\Delta$) evolves due to degradation only, neglecting the diffusion of the product. This is justified, since even for the fully degraded solution the polymer diffusion coefficient is two orders of magnitude smaller than that of the enzyme, see the~\SuppInfo.

By integrating Eq.~\ref{MM_for_mobility} over time, one obtains a mathematically elegant and physically insightful formulation of the space-dependent MM problem. As shown by Eq.~\ref{MM_equation_integral}, for any $z$ and $t$, the mobility $\Delta$ should depend only on $\mathcal{I}(z,t) = \int_{0}^{t}[E](z,t')\mathrm{d}t'$, the total amount of enzyme to which a portion of the substrate located in $z$ has been exposed, from the beginning of the experiment up to time $t$. To test whether the extension of the MM formalism to space-dependent conditions accounts for our data, we plot in Fig.~\ref{fig:5}A representative mobility data collected at various $z$, using the $\Delta$ \textit{vs} $\mathcal{I}$ representation suggested by 
Eq.~\ref{MM_equation_integral}. $\Delta(z,t)$ is obtained directly from the PCI analysis, while $\mathcal{I}$ is calculated by numerically integrating the fits to the experimental fluorescence profiles. Keeping in mind that $\mathcal{I}$ is initially zero and increases monotonically with time, we find that for all $z$ the mobility is initially zero, consistent with the notion that no degradation occurs until the substrate starts being exposed to the enzyme. As the cumulative exposure to the enzyme increases, so does the mobility. The plateau regime seen in Fig.~\ref{fig:5}A at large $\mathcal{I}$ corresponds to the largest mobility that our experiments can resolve
. While the overall shape of $\Delta$ \textit{vs} $\mathcal{I}$
is consistent with expectations, the data clearly vary systematically with $z$, whereas Eq.~\ref{MM_equation_integral} predicts a unique master curve, with a shape independent of $z$ and dictated only by the kinetic constants of the enzymatic reaction and the parameters describing the affine relation between mobility and products concentration.

Remarkably, the same features are also found for the other enzyme concentrations, with similar S-shaped curves exhibiting a systematic shift to the right (larger $\mathcal{I}$) as $z$ increases, similarly to the data sets shown in Fig.~\ref{fig:5}A, see Fig.~S9 in~\SuppInfo. This indicates that, as the enzyme diffuses deeper in the substrate ($z$ increases), a higher cumulative amount of enzyme (larger $\mathcal{I}$) is needed to reach a given degree of degradation. Thus, our experiments demonstrate unambiguously that the enzyme's efficiency in degrading the biopolymer decreases during the degradation process. 


The decrease in enzyme activity during degradation may stem from several causes. We check that the enzyme is stable 
in the absence of substrate: its specific activity, measured under reference assay conditions after incubation in the buffer, remains unchanged for at least 161 h, longer than the duration of the Fluo-PCI experiments (see~\SuppInfo). This rules out a spontaneous loss of activity over the experimental timescale, suggesting instead an inhibition mechanism resulting from the interaction with the substrate. In simple, or linear, inhibition, an inhibitor reduces the enzyme activity through a 1:1 stoichiometric reaction. This results in a MM-like degradation kinetics, albeit with modified apparent reaction constants~\cite{marangoni2003}, which is, however, incompatible with the systematic deviations with depth $z$ observed in our experiments. Instead, the observed progressive deactivation could stem from a more complex, substrate-dependent inactivation~\cite{sadana1988, ohs2018, hanson_number_2021}. 

In general, the decrease of enzyme activity may be modeled by introducing a time-dependent ``efficiency function'' $f(t)$, where $f(0)=1$ and $f$ decreases with time. When enzyme deactivation proceeds through multiple kinetic pathways or involves enzyme populations with different characteristic lifetimes, $f(t)$ can be phenomenologically described by a double exponential decay:
\begin{equation}
   f(t) = \chi \exp{\left [-k_{\alpha}t \right ]}+(1-\chi)\exp{\left[-k_{\beta}t \right]},
    \label{eq:double_exp}
\end{equation}
where $0 < \chi < 1$ is the relative weight of the first exponential relaxation in the right hand side of Eq.~(\ref{eq:double_exp}) and $k_{\alpha}$, $k_{\beta}$ represent apparent first-order rate constants obtained from a suitable combination of the true reaction constants involved in the specific deactivation pathway~\cite{aymard2000}. Although a molecular-level elucidation of the inactivation mechanism in our system remains to be established, following Ref.~\cite{aymard2000} we use  Eq.~(\ref{eq:double_exp}) as a phenomenological expression to model our data. We thus replace $[E]$ by $f(t)[E]$, the efficiency-corrected enzyme concentration, obtaining the following differential equation ruling the spatio-temporal evolution of the mobility:
\begin{equation}
    \frac{\mathrm{d}\Delta(z,t)}{\mathrm{d}t}=(k_{cat}/A)[E](z,t)f(t)\frac{\mathcal{C} - \Delta(z,t)}{(K_m/A)+\mathcal{C} - \Delta(z,t)} \,.
    \label{eq:MM_for_mobility_modified}
\end{equation}
Here, $A$ and $B$ are the coefficients in the affine relation between $\Delta$ and $[P]$, as in Eq.~\ref{MM_equation_integral}, $\mathcal{C}=([S]_0-B)/A$, $f(t)$ is given by Eq.~(\ref{eq:double_exp}), and $k_{cat}$ and $K_m$ are the catalytic rate and MM constant, respectively. We fit the model to the mobility data by solving numerically Eq.~\ref{eq:MM_for_mobility_modified} 
with $[E](z,t)$ obtained from the experimental evolution of the fluorescence signal. The fitting parameters are $k_{cat}/A$, $K_m/A$, $\mathcal{C}$, and $\chi$, $k_{\alpha}$, and $k_{\beta}$ introduced in the definition of $f(t)$. 
The fit is performed simultaneously on the 63 time-dependent mobility datasets, measured at 21 different depth values, $2.92\unit{mm} \le z \le 8.00\unit{mm}$, for each of the three experiments performed with biopolymer concentration $c = 20 \unit{g/L}$ and various enzyme concentrations. 
Figure~\ref{fig:2}A shows representative fits for three ROIs (lines), displaying a very good agreement with the data. From the global fit, the (rescaled) catalytic rate and MM constant are $k_{cat}/A = 1.33 \unit{(mM s)^{-1}}$ and $K_m/A = 1.34$, respectively, while the apparent rate constants for the enzyme inactivation are $k_{\alpha} = 4.6 \times 10^{-5} \unit{s^{-1}}$ and $k_{\beta} = 2.7 \times 10^{-6} \unit{s^{-1}}$, with $\chi = 0.92$.

In analogy to the simplified model not accounting for deactivation, 
it is convenient to integrate Eq.~(\ref{eq:MM_for_mobility_modified}) over time, in order to derive a master curve that should hold at all times and depths in the sample. The resulting expression is 
\begin{equation}
    \Delta(z,t)-\frac{K_m}{A} \ln{\left ( \frac{\mathcal{C}-\Delta(z,t)}{\mathcal{C}} \right)}=\frac{k_{cat}}{A} \Phi(z,t) \,.
    \label{MM_equation_integral_PHI}
\end{equation}
Here, the variable $\mathcal{I}$ introduced in Eq.~(\ref{MM_equation_integral}) is replaced by a new quantity $\Phi(z,t)$, proportional to the efficiency-weighted amount of enzyme to which a portion of substrate located in $z$ has been exposed, up to time $t$: $\Phi(z,t) = \int_0^t [E](z,t')f(t')\mathrm{d}t'$.  Equation~(\ref{MM_equation_integral_PHI}) defines a $\Delta(\Phi)$ master curve, whose shape is fixed once the fitting parameters are set, and which should hold for all $z$. We test this prediction for the experiment with $c = 20 \unit{g/L}$ and $[E]_{\infty} = 2.2 \unit{\mu M}$ in Fig.~\ref{fig:5}B, finding a very good overlap between data at different depths, and a good agreement with the model, shown by the yellow dashed line. Small deviations are seen, mostly for the ROIs at the top of the sample, likely due to some unavoidable mixing when gently pouring the enzyme solution on top of the polymer solution. Importantly, the good agreement between the model and the data is also seen for the other samples at the same $c$ but different $[E]_{\infty}$ (inset of Fig.~\ref{fig:5}B and Fig.~S9D-F in the~\SuppInfo), in contrast to the simpler model neglecting the decrease of enzyme activity. We further validate the double-exponential model accounting for the decrease of active enzymes, by showing that it faithfully accounts for the increase of product with time for long-term homogeneous kinetics data, Fig.~S3A in the~\SuppInfo, using the same $k_{\alpha}$, $k_{\beta}$, and $\chi = 0.92$ parameters as in Fig.~\ref{fig:5}B. 
Finally, the model also accounts for the mobility \textit{vs} $\Phi$ data for the mutant enzyme, see Fig. S12A,B of the~\SuppInfo. In this case, however, we find that $f(t)$ may be satisfactorily modeled by a single exponential decay.

The model also provides insight into the power law behavior of the degradation front reported in Fig.~\ref{fig:3}. With the parameters extracted from the fits to the experimental $\Delta(t,z)$, we compute mobility kymographs $\Delta_{calc} (t,z)$, shown in Fig. S10 of the~\SuppInfo, and retrieve the degradation fronts using the same Otsu algorithm as for the experiments. The degradation fronts $z^*(t^*)$ issued from the model are shown as lines in Fig.~\ref{fig:5}C, exhibiting an overall good agreement with the experimental data (symbols). This agreement demonstrates that the $z^* \sim {t^{*}}^{0.5}$ behavior observed experimentally and discussed in reference to Fig.~\ref{fig:3} is not the signature of diffusive behavior, but rather results from a complex interplay between enzyme diffusion,  enzymatic reaction, and decrease of enzyme activity. Our model fully captures this interplay, by introducing two key ingredients: the cumulative effect of enzymatic degradation, and the time-decaying efficiency of the enzyme as the substrate is degraded. Indeed, if the enzyme activity were assumed to be constant, the modeled degradation front would strongly deviate from the experimental one, exhibiting a much steeper $t^*$ dependence, see Fig.~S11 of the~\SuppInfo. The trend towards a sharper increase of $z^*$ in the absence of deactivation rationalizes the steeper slope of $z^*(t^*)$ for the mutant enzyme (stars in Fig.~\ref{fig:3}), suggesting that in our experiments the mutant form of the enzyme is more stable than the WT one. This is confirmed by the modeling, since for the mutant the decay rate of $f(t)$ is $1.03 \times 10^{-5}\unit{s^{-1}}$, almost 4.5 times slower than the rate $k_{\alpha} = 4.6 \times 10^{-5} \unit{s^{-1}}$ that accounts for 92\% of the efficiency loss of the WT enzyme.



\section*{Conclusion}\label{sec:cl}






The Fluo-PCI setup has enabled the simultaneous tracking of enzymatic degradation of biopolymers and diffusion of fluorescently labeled enzymes within the biopolymer matrix. We found that enzyme diffusion is independent of both enzymatic activity and biopolymer concentration, demonstrating that in our system enzyme mobility is decoupled from catalytic function, with no boost effect. By coupling real-time measurements of the substrate dynamics and enzyme diffusion, we fully resolved the kinetics of substrate degradation under realistic, spatially heterogeneous conditions. 

By extending the Michaelis-Menten formalism to spatially heterogeneous conditions and time-varying enzymatic activity, we proposed a simple model that fully rationalizes data for various substrate and enzyme concentrations and both wild type and mutant enzymes. For our system, the modeling unveiled a progressive loss of activity due to the interaction with the substrate. This process was accounted for by a double exponential decay of enzyme efficiency for the WT, and a simple exponential decay for the mutant. The origin of these different behaviors remains to be elucidated and is beyond the scope of the present study. Nevertheless, it raises an interesting  question of enzymology regarding how the alteration of the catalytic site may affect the long-term activity of an enzyme in the presence of its substrate. 

On a more general level, we believe that the experimental and modeling tools developed in this work will be highly valuable in real-world applications involving large and dense substrates and extended degradation times, where enzyme diffusion in a complex environment, enzyme longevity, and reaction kinetics dictate overall efficiency.

\section{Materials \& Methods}\label{sec:matmed}
\renewcommand{\theequation}{MM\arabic{equation}}
\setcounter{equation}{0}
\subsection*{Sample preparation}\label{subsec:sample}
Table I recapitulates the composition of the samples investigated in this work. The biopolymer is Feruloylated Arabinoxylan (FAX), which belongs to a class of hemicellulosic polysaccharides found primarily in the cell walls of grasses. FAX molecules are composed of a xylan backbone substituted with arabinose residues, some of which are linked to a ferulic acid (FA) through an ester bond. FAX has been extracted from wheat endosperm as a powder with $63$ \% purity~\cite{carvajal2005}, all other chemicals have been purchased from Sigma-Aldrich. The molecular weight of FAX is $M_w \simeq 4.4 \times 10^5 \unit{Da}$, the arabinose/xylose  substitution ratio is $0.58$, the total FA content is $1.68 \unit{\mu g}$ per mg of FAX , and its specific volume is $0.596 \unit{cm^3/g}$. 
The FAX powder is stored in a desiccator (P\textsubscript{2}O\textsubscript{5} atmosphere) at room temperature. Batches of FAX solution are prepared by dissolving the FAX powder in a $50$ mM citrate phosphate buffer (pH~$5$) containing $0.02 \unit{wt \%}$ Thiomersal as a preservative, to a final concentration of $c = 20 \unit{g/L}$. The mixture is stirred vigorously for $24$ hours at room temperature. The solution is then separated into different aliquots and further diluted to $c = 10 \unit{g/L}$ when needed. The aliquots are stored at $4 \celsius$ for at least $48$ hours before use. FAX chains are linear neutral chains that behave in the buffer as a semiflexible polymer in good solvent conditions~\cite{carvajal2005}.
Considering that the FAX solutions are in the semidilute regime in the investigated concentration range ($c^* \simeq 2 \unit{g/L}$)~\cite{carvajal2005}, the characteristic mesh size~\cite{rubinstein2003} $\xi$ of the FAX visco-elastic solution is $\xi \sim l_p \phi^{-\nu/(3\nu-1)}$, where $\nu = 0.588$ is the Flory exponent, $l_p \simeq 5 \unit{nm}$ is the typical persistence length of arabinoxylan, corresponding to approximately $8$ xylose units~\cite{petermann2023} as measured by scattering techniques~\cite{petermann2023, yu2018}, and $\phi$ is the monomer volume fraction. We estimate  $\xi \simeq 45 \unit{nm}$ at $c = 10 \unit{g/L}$ and  $\xi \simeq 30 \unit{nm}$ at $c = 20 \unit{g/L}$, much larger than the enzyme size ($R_s \simeq 3 \unit{nm}$).

The enzyme used in this study, \textit{Np}Xyn11A, randomly cleaves the X-X bonds along the FAX backbone~\cite{vardakou2008}, excluding however the X-X bonds in which at least one of the xylose residues is substituted. The wild-type form of \textit{Np}Xyn11A (WT) corresponds to the native enzyme sequence. The mutant enzyme corresponds to a E113A variant of \textit{Np}Xyn11A, in which the glutamic acid residue at position 113, acting as the catalytic nucleophile, is replaced by alanine. This mutation strongly reduces the enzymatic activity compared to the WT, as assessed by specific-activity measurements on arabinoxylan under standard assay conditions (see~\SuppInfo). Both the WT and the mutant enzymes have a theoretical molar mass of $\sim 26 \unit{kDa}$ and a roughly spherical 3D structure, with a geometrical diameter estimated to $4 \unit{nm}$ using PyMOL~\cite{PyMOL}. 

The enzymes are recombinantly expressed in \textit{E. coli} and purified as described in the~\SuppInfo. The purified enzymes are then labeled with tetramethylrhodamine-5-(and-6)-isothiocyanate (TRITC), which has excitation and emission maxima at 555 and 580 nm, respectively. The labeling protocol is detailed in the~\SuppInfo; in brief, TRITC reacts covalently with lysine residues of the enzymes, leading to an average labeling degree of 0.9–1.05  TRITC molecules per enzyme molecule, as determined by spectrophotometry. The protocol includes a purification step to remove unbound fluorophore. Immediately after labeling and purification, unbound TRITC is estimated to account for approximately 20\% of the total fluorescence intensity. This fraction increases to about 50\% in the samples used for the Fluo-PCI experiments (see Experimental Results and~\SuppInfo). This increase is most likely due to the progressive release of protein-bound TRITC over the 7-9 months separating enzyme production from the Fluo-PCI experiments, during which the enzymes are stored at -20°C (see~\SuppInfo).
\\ 

\begin{table}
    \centering
    \begin{tabular}{c c c c c}
    \hline
    $c$ [g/L] && Enzyme type && $[E]_{\infty}$ [$\mu$M]\\
    \hline
    20 && \textbf{WT} && 0.55  \\
    20 && \textbf{WT}  && 1.1  \\
    20 && \textbf{WT}  && 2.2  \\
    10 && \textbf{WT}   &&  2.2\\
    10 && \textbf{mutant}  && 2.2 \\
    \hline
    \end{tabular}
    \caption{List of experimental conditions. $c$ is the biopolymer (FAX) concentration, $[E]_{\infty}$ is the enzyme (xylanase) concentration averaged over the sample volume, WT stands for wild-type enzyme. }
    \label{table:samples}
\end{table}

\subsection*{Experimental Setup}
A collimated laser beam (Verdi V-2 by Coherent, in-vacuo wavelength $\lambda_0 = 532.5 \unit{nm}$ and maximum power $2 \unit{W}$) is expanded in the vertical direction by a Powell lens with divergence $2\phi = 30^{\circ}$ (Thorlabs LGL130) and collimated by a cylindrical lens (focal length $25.2 \unit{mm}$) to form a laser sheet of height $\approx 12 \unit{mm}$ and thickness $\approx 1.5 \unit{mm}$. The laser sheet impinges on the sample cuvette, immersed in a transparent water bath to control temperature. For all experiments, temperature is fixed at  $T = 37.0 \pm 0.2 \celsius$.  The scattered light is separated by a shortpass dichroic mirror (Thorlabs DMSP550) from the fluorescence emission ($\lambda_{em} = 580$ \unit{nm}), which is further filtered before camera 2 by an interference emission filter (Thorlabs MF620-52). The imaging lenses of both the scattering and the fluorescence paths have focal length $150 \unit{mm}$. In the scattering path, a diaphragm placed in the focal plane of the imaging lens allows for a precise selection of the scattering wave vector $q = 4 \pi n \sin (\theta/2)/\lambda_0 \simeq 22.2 \unit{\mu m^{-1}}$, with $n = 1.3315$ the index of refraction of  water, and $\theta=90$ deg. The scattering cuvette (FisherBrand semi-micro, rectangular cross section $4$ mm $\times$ $10$ mm and height $40$ mm) is filled with $420 \unit{\mu L}$ of FAX solution and immersed in the thermostated bath, approximately $1 \unit{hour}$ before starting the measurements. $70 \unit{\mu L}$ of labeled enzyme solution is then gently poured on top of the FAX solution using an Eppendorf micropipette. The cuvette is then sealed with a UV-glued homemade lid that ensures negligible evaporation throughout the experiment.

\subsection*{Computation of polymer mobility and degradation front}\label{subsec:mobility}
The microscopic dynamics of the polymer solution at a given $z$ is measured by the degree of correlation  $c_I(z, t,\tau)$ of the scattered intensity between times $t$ and $t+\tau$:
\begin{equation}
    c_I(z, t,\tau) = \dfrac{\langle I_p(t)I_p(t+\tau) \rangle_{z}}{\langle I_p(t) \rangle_{z} \langle I_p(t+\tau) \rangle_{z}}-1,
    \label{eq:ci}
\end{equation}
where $I_p$ is the value of the scattered intensity detected by the $p$-th pixel of the camera and $\langle \dots \rangle_z$ denotes the spatial average over a Region of Interest (ROI) centered at a height $z$, with size $568 \times 40 \unit{pixels}$, corresponding to $3.607 \unit{mm} \times 0.254 \unit{mm}$ in the sample. The degree of correlation $c_I$
is a space-and time-resolved equivalent of the intensity correlation function $g_2(\tau)-1$ measured in conventional dynamic light scattering~\cite{berne2000}:
\begin{equation}
    g_2(\tau) - 1 = \left < \left < c_I(z, t,\tau) \right >_{ROIs} \right >_t ,
    \label{eq:g2}
\end{equation}
where the $ROIs$ and $t$ subscripts indicate averages over the full scattering volume and time, respectively. Both $c_I$ and $g_2-1$ quantify the microscopic dynamics on a length scale of order $1/q \approx 50 \unit{nm}$. To track the evolution of the polymer undergoing degradation, we define a dimensionless mobility $\Delta (z,t)$:
\begin{equation}
\Delta (z,t) = -\ln{c_I (z,t, \tau = \Bar{\tau})}.
    \label{Eq:chain mobility}
\end{equation}
For scatterers undergoing Brownian motion with diffusion coefficient $D$, $\Delta(t,z) = (q^2 \langle r^2 \rangle/3)$ is proportional to the mean square displacement $\langle r^2 \rangle^= 6D \Bar{\tau}$ over a time interval $\Bar{\tau}$~\cite{berne2000}. In our case $\Delta$ corresponds more generally to a biopolymer microscopic mobility, which increases as the polymer size is reduced by the action of the enzymes. We chose $\Bar{\tau} = 20 \unit{ms}$ because it allows us to follow with good temporal resolution the full evolution of the microscopic dynamics of the substrate, from the pristine viscoelastic state to a fluid, degraded state. As seen in Fig.~\ref{fig:2}A, for $\Delta \gtrsim 4$ (corresponding to $c_I \lesssim 0.01$) the microscopic mobility exhibits significant scatter: this is due to the experimental noise on approaching the theoretical $c_I=0$ baseline corresponding to motion on length-scales $>> 1/q$. 

The degradation front is identified by applying the Otsu thresholding algorithm~\cite{otsu1979}, which selects a threshold $\Delta^*$ that minimizes the intra-class variance of the two mobility populations seen in the histogram of the mobility map (Fig.~S4 
in the~\SuppInfo). The two populations correspond to low-mobility, non-degraded biopolymer and high-mobility, degraded biopolymer, respectively. The threshold $\Delta^*$ obtained from the Otsu algorithm defines the mobility value at the degradation front. For a given ROI (i.e., a given front position $z^*$), there are typically a few times $t$ such that $\Delta(t) = \Delta^*$, due to experimental uncertainties. We define the front time $t^*$ and its uncertainty $\delta t^*$ (error bars in Fig.~\ref{fig:3}) as the average and standard deviation, respectively, of these times. The uncertainty on $z^*$ is set to $127 \unit{\mu m}$, half the size along $z$ of the ROIs. Finally, the errors bars in Fig.~\ref{fig:2}C are the standard deviation of $\Delta(t,z^*)$ for $t\in [t^*- \delta t^*,t^* + \delta t^*]$.

\subsection*{Fluorescence signal modeling and enzyme diffusion}\label{subsec:fluorescence}
We model the time evolution of the fluorescence intensity profiles $I(t;z)$ using the analytical solution of the 1D diffusion equation~\cite{crank1979}, since all quantities depend only on depth $z$ and time $t$. We assume that the signal is due to the labeled enzyme and to free fluorophore, each contributing an intensity level proportional to its local concentration: 
\begin{equation}
I(t;z) = rI_l(t;z)+[1-r]I_s(t;z),
\label{eq:diffusion_solution}
\end{equation}
where the subscripts $l$ and $s$ refer to the larger and smaller species, e.g. the labeled enzyme and the free fluorophore, respectively, and $r$ is the intensity-weighted contribution of the enzymes, relative to the total intensity. The boundary conditions are impermeable top and bottom walls (zero flux), and the initial conditions are
\begin{equation*}
    \begin{cases}
    I_{i}(0;z) = I_{0,i}, \quad -z_0<z<0\\
    I_{i}(0;z) = 0, \, \,\, \qquad 0<z<L
    \end{cases}
\end{equation*}
with $i = l,s$, $L \simeq 10.5 \unit{mm}$ the initial height of the polymer solution, and $z_0 \simeq 1.75 \unit{mm}$ the initial height of the enzyme solution deposited on top of the polymer solution. With these boundary and initial conditions, the solution to the 1D diffusion is
~\cite{crank1979} 
\begin{align}
I_{i}(t;z)
&= I_{0,i}
\frac{2}{\pi}
\sum_{m=1}^{\infty}
\frac{1}{m}
\sin\!\left(\frac{\pi m z_0}{L+z_0}\right)
\cos\!\left(\frac{\pi m z}{L+z_0}\right) \nonumber \\
&\quad\times \exp\!\left[-\frac{(\pi m)^2 D_{i} t}{(L+z_0)^2}\right]
+ I_{0,i}\frac{z_0}{L+z_0}
,
\quad i = l,s  \label{Eq:Cranck1}
\end{align}
with $D_l$ and $D_s$ the diffusion coefficients of the enzymes and free fluorophores, respectively.
We use Eqs.~(\ref{eq:diffusion_solution}) and (\ref{Eq:Cranck1}) to fit simultaneously the $I(t;z)$ data for all ROIs with $z \ge 2.92 \unit{mm}$, for all samples listed in Table~\ref{table:samples}, finding the parameters $r, D_l $ and $D_s$  reported in the main text.

To confirm that the coexistence of two independent diffusing species is needed to satisfactorily fit our data, we compare the goodness of fit of this model with that of a single-species diffusion model, using the Bayesian Information Criterion ($BIC$) parameter, $BIC = n\ln{(RSS/n)}+p\ln{(n)}$. Here  $n$ is the number of data points, $p$ is the number of free parameters and $RSS= \sum_i r_i^2$ is the sum of the squared residuals $r_i^2$~\cite{kass1995}. 
Better fits correspond to lower values of $BIC$. The $BIC$ parameter obtained from a global fit with one diffusing species is $BIC_1 \simeq 6.3 \times 10^5$, while for the global fit with two diffusing species $BIC_2 \simeq 3.5 \times 10^4$, one order of magnitude smaller than $BIC_1$. To further confirm the value $r=0.47\pm 0.03$ issued from the global fit, we perform additional fits imposing fixed values of $r$. Figure~S6 of the~\SuppInfo~shows the $r$ dependence of the root mean square error $RMSE = \sqrt{RSS/n}$ of the two-species model. For the global fit, we indeed find that $RMSE$ is minimum for $r=0.47$ (gray hexagons in Fig.~S6); a similar behavior is obtained also when fitting one by one the data sets of different experiments (colored symbols in Fig.~S6
).

We confirm these results by independently measuring the diffusion coefficient of the purified, labeled enzyme in the buffer solution using Taylor Dispersion Analysis (TDA), a technique particularly well suited for determining the size of sub-nanometric and nanometric species in solution~\cite{cipelletti2015}. The TDA measurements clearly reveal the presence of two distinct species in solution, with parameters consistent with those derived from the fits of $I(t;z)$, see Fig.~S5 and the~\SuppInfo~for details. This result is further corroborated by gel electrophoresis measurements performed on the fluorescently labeled enzymes after the degradation experiments, as described in the~\SuppInfo~(Fig. S2)

\subsection*{Michaelis-Menten model in heterogeneous degradation conditions}
The Michaelis-Menten (MM) model~\cite{marangoni2003} describes the enzymatic reaction in two steps. The enzyme E reversibly binds with rate constant $k_1$ and unbinds with rate constant $k_{-1}$ to its substrate S, forming a substrate-enzyme complex ES, which breaks down at a catalytic rate $k_{cat}$, releasing the reaction product P and the enzyme E: 
\begin{equation}
    E + S \;\overset{k_1}{\underset{k_{-1}}{\rightleftharpoons}}\; ES
\;\xrightarrow{k_{cat}}\; E + P
    \label{MM_mechanism}
\end{equation}

Our experiments are run with a large excess of substrate with respect to the enzyme $[S] \gg [E]$, which implies that the concentration of the substrate-enzyme complex is constant throughout the reaction (steady-state approximation). Under these conditions and taking into account that in our experiments the concentrations of both enzyme and product depend not only on time, but also on depth $z$ in the sample, Eq.~(\ref{MM_mechanism}) leads to a MM-like differential equation for the evolution of the product concentration $[P]$:
\begin{equation}
    \frac{\mathrm{d}[P] (z,t)}{\mathrm{d}t}=k_{cat}[E](z,t)\frac{[S]_0 - [P](z,t)}{K_m+[S]_0 - [P](z,t)}
    \label{MM_equation_classic}
\end{equation}
Here $K_m = (k_{-1} + k_{cat})/{k_1}$ is the MM constant and $[S]_0$ is the initial substrate concentration. 

We establish a direct connection between $[P]$ in Eq.~(\ref{MM_equation_classic}) and the mobility $\Delta$ measured in our experiments through long-term kinetics measurements in homogeneous conditions (i.e., constant and uniform concentration of enzyme in a biopolymer solution), finding  $[P](t) = A \Delta(t) + B$, where $[P](t)$ is measured through dinitrosalicylic acid assay and $\Delta(t)$ is measured by dynamic light scattering (see Fig.~S3 
and related text in~\SuppInfo~for details). Inserting this relation in Eq.~(\ref{MM_equation_classic}), one obtains a differential equation for $\Delta (t,z)$: 
\begin{equation}
    \frac{\mathrm{d}\Delta (z,t)}{\mathrm{d}t}=(k_{cat}/A)[E](z, t)\frac{\mathcal{C} - \Delta (z,t)}{(K_m/A)+\mathcal{C} - \Delta(z,t)}
    \label{MM_for_mobility}
\end{equation}
Equation~(\ref{MM_for_mobility}) is a MM-like equation for the polymer mobility, of the same form as Eq.~(\ref{MM_equation_classic}), but with ``apparent'' kinetic parameters $k_{cat}/A$, $K_m/A$ and $\mathcal{C}=([S]_0-B)/A$. By integrating over time Eq.~(\ref{MM_for_mobility}), one obtains
\begin{equation}
    \Delta(z,t)-(K_m/A) \ln{\left ( \frac{\mathcal{C}-\Delta(z,t)}{\mathcal{C}} \right)}=(k_{cat}/A)\mathcal{I}(z,t) \,,
    \label{MM_equation_integral}
\end{equation}
which shows that the mobility spatio-temporal profiles should only depend on the quantity $\mathcal{I} = \int_0^t [E](z,t')\mathrm{d}t'$. As discussed in the main text, a more advanced model, accounting for enzyme deactivation, is in fact required to reproduce our experiments.
%

\section*{Data Availability}
Data have been deposited in Zenodo (DOI: to be inserted later).

\section*{Author Contributions}
Author contribution: \textbf{Vincenzo Ruzzi}: Conceptualization, Methodology, Software, Validation, Formal analysis, Investigation, Data Curation, Writing - Original Draft, Writing - Review \& Editing, Visualization. \textbf{Antoine Bouchoux}: Conceptualization, Methodology, Validation, Resources, Writing - Original Draft, Writing - Review \& Editing, Supervision, Project administration, Funding acquisition. \textbf{Carole Antoine-Assor}: Conceptualization, Validation, Investigation, Resources, Writing - Review \& Editing, Supervision, Project administration, Funding acquisition. \textbf{Donna-Joe Bigot}: Methodology, Formal analysis, Investigation. \textbf{Salma Menzeh}: Methodology, Formal analysis, Investigation. \textbf{Maike Petermann}: Methodology, Formal analysis, Investigation. \textbf{Laurent Leclercq}: Methodology, Formal analysis, Investigation. \textbf{Herv\'e Cottet}: Methodology, Formal analysis, Supervision. \textbf{C\'edric Montanier}: Conceptualization, Methodology, Validation, Resources, Supervision, Project administration, Funding acquisition. \textbf{Claire Dumon}:  Conceptualization, Methodology, Validation, Resources, Writing - Review \& Editing, Supervision, Project administration, Funding acquisition. \textbf{Luca Cipelletti}: Conceptualization, Methodology, Software, Validation, Resources, Formal analysis, Data Curation, Writing - Original Draft, Writing - Review \& Editing, Visualization, Supervision, Project administration, Funding acquisition. \textbf{Laurence Ramos}: Conceptualization, Methodology, Validation, Resources, Formal analysis, Data Curation, Writing - Original Draft, Writing - Review \& Editing, Visualization, Supervision, Project administration, Funding acquisition.

\begin{acknowledgments}
We gratefully acknowledge W. Poon and R. Merindol for useful discussions, and M. Milani for preliminary measurements. A. Chopin is thanked for help in the analysis of TDA data. This work received government funding administered by the French National Research Agency (ANR) under the France 2030 initiative (ANR-24-RRII-0003, EXPLOR'AE program led by INRAE). We acknowledge also financial support from the French Agence Nationale de la Recherche (ANR) (Grant No. ANR-19-CE06-0030-02, BOGUS, and Grant No. ANR-25-CE06-1825, CATCHY), and from Labex Numev (AAP2022-1-12-Ramos).
\end{acknowledgments}

\bibliography{bibliography_Catchy1}

\appendix
\renewcommand{\thefigure}{S\arabic{figure}}
\setcounter{figure}{0}
\renewcommand{\thetable}{S\arabic{table}}
\setcounter{table}{0}
\renewcommand{\theequation}{S\arabic{equation}}
\setcounter{equation}{0}
\newpage
\onecolumngrid
\vspace{2em}
\begin{center}
  \rule{0.8\textwidth}{0.4pt}\\[0.5em]
  {\large\textbf{Supplementary Information}}\\[0.3em]
  {\normalsize Simultaneous measurements of enzyme propagation and biopolymer mobility
  elucidate the directional degradation of a dense biopolymer matrix}\\[0.3em]
  {\small Vincenzo Ruzzi \textit{et al.}}\\[0.3em]
  \rule{0.8\textwidth}{0.4pt}
\end{center}

\section*{Setup and acquisition details}

To minimize the number of acquired images, camera 1 (CMOS Hamamatsu Orca-Flash V3) is triggered by an external TTL signal with a variable frame rate scheme as described elsewhere~\cite{philippe2016}, with a minimum delay of  $20 \unit{ms}$, an average acquisition rate of $1$ fps and an exposure time of $1$ ms. The fluorescence camera (camera 2, CMOS acA2000 340km, Basler) works at a constant frame rate of $0.05$ fps and an exposure time of $150$ ms. The total acquisition time ranges from $24$ to $60$ hours. To avoid possible photo-bleaching effects because of long sample exposure to the incident laser light, we use a mechanical shutter placed along the incident light path (see Fig.~1 in the main paper), which blocks the laser beam while the cameras are not acquiring an image.

Given the geometry of the experiment (the enzyme solution is poured on top of the biopolymer solution), the directional degradation of the biopolymer and the propagation of the enzyme are investigated as a function of the $z$ coordinate along the vertical direction. Thus, the fluorescence and speckle images (field of view $\sim 3.7  \unit{mm} \times 9.0  \unit{mm}$) are subdivided into Regions of Interests (ROIs) of size $3.607 \unit{mm} \times 0.254 \unit{mm}$, stacked in the vertical direction. From the time series of the images, we extract the temporal evolution of the fluorescence intensity, $I$, and of the biopolymer mobility, $\Delta$, see Eqs.~(MM4) and (MM3) in Sec. \MM in the main paper, respectively. The typical size and position of the ROIs are reported in Fig.~1B in the main paper. 

\section*{Characterization of pristine and degraded biopolymer solutions}

\begin{figure}[t!]
    \centering
    \includegraphics[width=0.5\columnwidth]{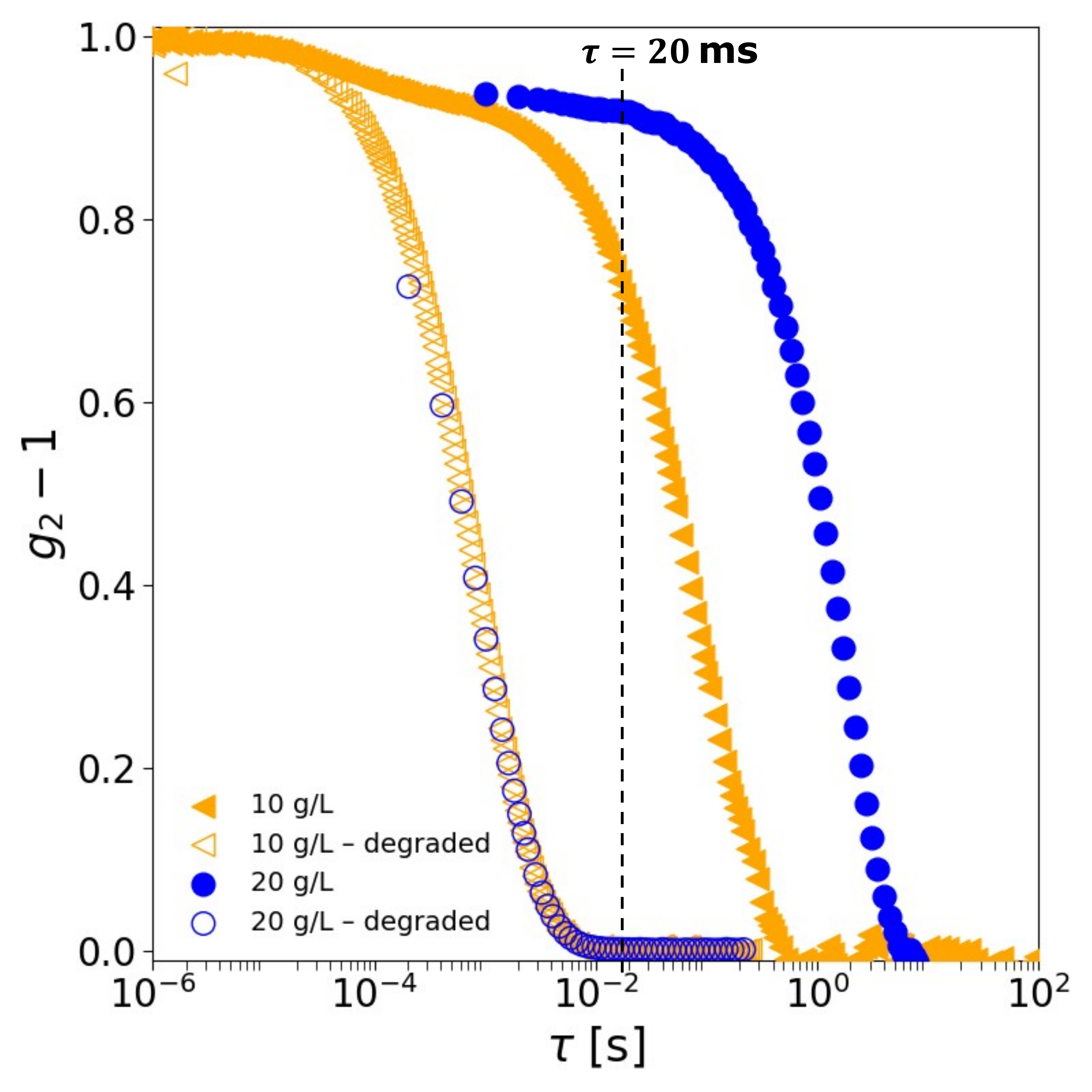}
    \caption{Intensity correlation functions, $g_2-1$, of the pristine FAX solutions (full symbols) and after complete degradation (open symbols). The dashed line indicates the lag-time $\Bar{\tau} = 20 \unit{ms}$ at which the biopolymer mobility $\Delta$ is evaluated.}
    \label{fig:S6}
\end{figure}

We measure by dynamic light scattering (DLS~\cite{berne2000}) the intensity correlation functions $g_2(\tau)-1$ for pristine biopolymer solutions and solutions of degraded biopolymer, with concentration $c = 10 \unit{g/L}$ and $20 \unit{g/L}$. The intensity correlation functions $g_2(\tau)-1$ in Fig.~\ref{fig:S6} have been measured either by the camera-based Photon Correlation Imaging (PCI) setup ($c = 20 \unit{g/L}$), as described in the main paper, or by standard DLS ($c = 10 \unit{g/L}$), at $T = 37 \celsius$ and scattering wave vector $q = 22.2 \unit{\mu m^{-1}}$. For the PCI data, the intensity correlation function is computed from the temporal average of the degree of correlation over a suitable time interval $\Delta t$: $g_2(\tau)-1 = \langle {c_I(t,\tau)} \rangle_{\Delta t}$. The DLS setup is a LS Instrument Spectrometer\textsuperscript{\texttrademark}, equipped with a laser operating at $\lambda_0 = 632.8 \unit{nm}$ and a goniometer enabling measurements at scattering angles $\theta \in [20^{\circ}, 140^{\circ}]$.  

The overlap of the correlation functions of the solutions of degraded biopolymer at $10 \unit{g/L}$ (orange triangles) and $20 \unit{g/L}$ (blue circles) indicates that the microscopic dynamics for the final degraded state is the same regardless of the initial biopolymer concentration. This suggests that the reduction of molecular weight of the FAX chains, which leads to an increase of the overlap concentration $c^*$~\cite{rubinstein2003}, is such that the solutions of degraded biopolymer are in dilute conditions, where the dynamics are independent of concentration. For these degraded biopolymer solutions, the relaxation time of $g_2-1$ is $\tau_r \approx 1 \unit{ms}$, corresponding to an effective polymer diffusion coefficient~\cite{berne2000} $D_{eff} = 1/(2q^2 \tau_r) \approx 10^{-12} \unit{m^2s^{-1}}$. As argued in the main text, $D_{eff}$ is about 100 times smaller than the diffusion coefficient of the enzyme,   justifying the simplifying assumption that polymer diffusion may be neglected with respect to that of the enzyme.

\section*{Enzyme production, labeling, activity and stability}

\subsection*{Enzyme production and purification}

\textit{Np}Xyn11A wild-type (WT) and mutant \textit{Np}Xyn11A E113A are expressed in \textit{Escherichia coli} strain BL21(DE3) as described elsewhere~\cite{vardakou2008}. Briefly, cells are cultured in Luria-Bertani broth at $37 \celsius$ until mid-exponential phase, corresponding to an optical density of $0.6$ at $600 \unit{nm}$. Protein expression is induced by the addition of isopropyl-$\beta$-D-thiogalactopyranoside to a final concentration of $1$ mM and further incubation for $4$ h at $37 \celsius$. Harvested cells are resuspended in $50$ mM phosphate buffer, pH $7.2$, containing $300$ mM NaCl and a protease inhibitor cocktail (SigmaFAST protease inhibitor cocktail, Sigma-Aldrich). Cells are lysed by sonication on ice for $1$ min, and the resulting lysate is clarified by centrifugation for $30$ min at $74 000 \times~g$ and $4 \celsius$. Proteins are purified by immobilized metal ion affinity chromatography (IMAC) using TALON\textsuperscript{\textregistered} Metal Affinity Resin ($2.5 \unit{mL}$ bed volume, Clontech), as described by Enjalbert \textit{et al.}~\cite{enjalbert2020}.  Briefly, proteins are eluted in $50$ mM sodium phosphate buffer, pH $7.2$, containing $300$ mM NaCl and $100$ mM imidazole. The buffer is then exchanged in a PD-10 desalting column (Cytiva)  against $50$ mM sodium phosphate buffer, pH $7.2$. Purified proteins are assessed as homogeneous by sodium dodecyl sulfate–polyacrylamide gel electrophoresis (SDS-PAGE) under denaturing and reducing conditions (Any kD\textsuperscript{\texttrademark} Mini-PROTEAN\textsuperscript{\textregistered} TGX Stain-Free\textsuperscript{\texttrademark} Protein Gels, Bio-Rad). Protein concentrations are determined by measuring the absorbance at $280$ nm and applying the Beer-Lambert law. The theoretical molar extinction coefficient is $61880 \unit{M^{-1} cm^{-1}}$ for both wild type and mutant enzymes. Final enzyme concentrations are about $19 - 22 \unit{\mu M}$. All experiments described in the main text and the Supplementary Information have been performed with one enzyme batch (see Table S1 and Table 1 in the main text), except for the long term thermal stability and the homogeneous degradation with biopolymer concentration  $c = 20 \unit{g/L}$ (see Fig.~\ref{fig:S7}D and related text).

\subsection*{Enzyme dye labeling}
\begin{figure}
    \centering
    \includegraphics[width=0.5\columnwidth]{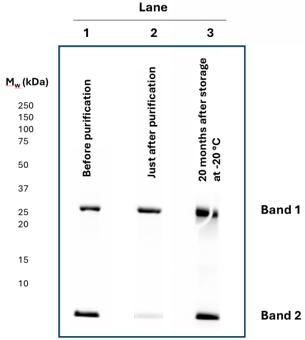}
    \caption{SDS-PAGE analysis of TRITC-labeled \textit{Np}Xyn11A WT during purification and storage. Fluorescence image of a SDS-PAGE gel showing the WT enzyme before purification (lane 1), just after purification (lane 2), and after 20 months of storage at -20 °C (lane 3). Band 1 corresponds to the TRITC-labeled enzyme, whereas band 2 corresponds to free, unbound TRITC.}
    \label{fig:SDS-page}
\end{figure}

The purified enzymes are labeled with tetramethylrhodamine-5-(and-6)-isothiocyanate (TRITC, Sigma-Aldrich) according to a protocol adapted from the manufacturer’s guidelines. First, $1$ mL of enzyme solution is prepared at approximately $2 \unit{\mu M}$ in 50 mM sodium phosphate buffer, pH $8.3$. TRITC, previously dissolved in $25 \unit{\mu L}$ of dimethyl sulfoxide (DMSO, Sigma-Aldrich), is rapidly added to the enzyme solution to reach a final TRITC concentration of $20 \unit{\mu M}$, corresponding to a $10$-fold molar excess of fluorophore over enzyme. The mixture is incubated at room temperature for $4$ h, followed by an additional incubation at $4 \celsius$ for $20$ h. Excess TRITC is removed by sequential affinity chromatography followed by desalting column. The labeling reaction is diluted 10-fold in $50 \unit{mM}$ sodium phosphate buffer, $300 \unit{mM}$ NaCl, pH 7.4, and applied to TALON\textsuperscript{\textregistered} metal affinity resin, as described above. The column is washed with $35 \unit{mL}$ of the dilution buffer followed by two steps of $5 \unit{mL}$ dilution buffer containing $5 \unit{mM}$ and $10 \unit{mM}$ imidazole, respectively. The solution of labeled protein is then eluted with $100 \unit{mM}$ imidazole. The eluate is dialyzed against $50 \unit{mM}$ sodium phosphate buffer (pH 7.4) and subsequently concentrated to approximately $3.5 \unit{mL}$ using $10$-kDa MWCO centrifugal filters. Residual free TRITC is further removed using a PD-10 column. The recovered protein fraction is then passed through a second PD-10 column under the same conditions. The labeled enzyme solutions are analyzed by SDS-PAGE under denaturing and reducing conditions (Any kD\textsuperscript{\texttrademark} Mini-PROTEAN\textsuperscript{\textregistered} TGX Stain-Free\textsuperscript{\texttrademark} Protein Gels, Bio-Rad). The gel is imaged by in-gel fluorescence using excitation at $550$ nm and an emission filter centered at $595$ nm, both selected to match the fluorescence properties of TRITC (see Fig.~\ref{fig:SDS-page} for the WT-labeled enzyme). To assess the removal of unbound fluorophore, we compare the intensity of the upper fluorescent band, corresponding to the labeled enzyme, to the lower band (lane 2 in Fig.~\ref{fig:SDS-page}), corresponding to free TRITC. Before purification, approximately $50$ \% of the total fluorescence intensity is attributed to free TRITC for WT and mutant enzymes (lane 1 in Fig.~\ref{fig:SDS-page}, shown for the WT enzyme). After the purification steps described above, this fraction decreases to approximately $20$ \% of the total fluorescence intensity (lane 2 in Fig.~\ref{fig:SDS-page}). However, the TDA analysis and the modeling of the fluorescence signal presented below indicate that this fraction increases again to about $50$ \% after $7-9$ months of storage at $-20 \celsius$, i.e. at the time of the PCI/fluorescence experiments. This increase is most likely due to the progressive release of protein-bound TRITC over time, possibly through destabilization of some thiourea bonds between the enzyme and the fluorophore under the storage conditions. Although no SDS-PAGE image is available for the enzyme solutions after $7-9$ months, SDS-PAGE analysis after $20$ months of storage at $-20 \celsius$ shows a much stronger free-TRITC signal than immediately after labeling and purification, again accounting for approximately 50 \% of the total fluorescence intensity (lane 3 in Fig.~\ref{fig:SDS-page}).

\subsection*{Enzyme activity}

Enzymatic activities are measured using the 3,5-dinitrosalicylic acid (DNS) assay~\cite{miller1959}. This colorimetric assay quantifies reducing sugar extremities through their reaction with DNS, which produces a colored compound detected by an absorbance measurement. 
As the enzyme cleaves the arabinoxylan backbone, new reducing ends are generated. Using a xylose calibration curve, the absorbance measured after enzymatic degradation is converted into a concentration of xylose equivalents, which reflects the amount of reducing ends produced during hydrolysis, after correction for the initial reducing-end content of the substrate. The degradation rate is then obtained from the increase in xylose-equivalent concentration over time. The specific activity is used as a measure of the catalytic efficiency of the enzyme toward a given substrate under defined physico-chemical conditions. It is expressed in $\unit{\mu mol \, min^{-1} mg^{-1}}$, or equivalently in $\unit{U \, mg^{-1}}$. In this convention, 1 U corresponds to the production of $1 \unit{\mu mol}$ of xylose equivalents per minute under the assay conditions. 

The specific activities reported in Table~\ref{table:activity} are measured under reference conditions for xylanase activity assays, using wheat arabinoxylan (WAX) at $10 \unit{g/L}$ in $50 \unit{mM}$ sodium phosphate buffer at pH $6$ and at $37 \celsius$. Under these reference conditions, the labeled WT enzyme exhibits a specific activity of $4520 \pm 307 \unit{U \, mg^{-1}}$, whereas that of the labeled E113A mutant is $1.520 \pm 0.021 \unit{U \, mg^{-1}}$, corresponding to an approximately $3000$-fold decrease in activity. These measurements therefore confirm the strongly reduced catalytic activity of the E113A mutant compared to that of the WT enzyme.

\begin{table}[t!]
    \centering
    \begin{tabular}{c c c c c}
    \hline
    Enzyme type && Specific activity [$\unit{U \, mg^{-1}}$]\\
    \hline
     \textbf{WT}, labeled && $4520 \pm 307$  \\
    \textbf{E113A mutant}, labeled  && $1.520 \pm 0.021$   \\
    \hline
    \end{tabular}
    \caption{Specific activities of the labeled wild type and mutant enzymes, as determined using the DNS assay.}
    \label{table:activity}
\end{table}

\subsection*{Enzyme stability}

To assess the stability of the WT enzyme and its catalytic activity over time in the absence of substrate, a $2.2$ $\mu$M enzyme solution is incubated under the buffer and temperature conditions used in the Fluo-PCI experiments, i.e., in $50$ mM citrate-phosphate buffer at pH $5$ and temperature $37^{\circ}$C. Aliquots are collected at different incubation times, up to $161$ h, and the specific activity of the enzyme is measured using the DNS assay described above. Activity measurements are performed under the reference conditions used for xylanase activity assays, i.e., using wheat arabinoxylan (WAX) at $10$ g/L in $50$ mM sodium phosphate buffer at pH $6$ and at $37^{\circ}$C. The specific activities measured after the different incubation times are reported in Table~\ref{table:stability}. Despite some variability, the activity remains of the same order of magnitude throughout the experiment and shows no systematic decrease with incubation time. In particular, no substantial progressive loss of activity is observed over the timescale of the Fluo-PCI experiments (up to $60$ h). These results therefore indicate that the time-dependent decrease in enzymatic activity inferred from the heterogeneous degradation experiments cannot be attributed to spontaneous enzyme inactivation in buffer due to prolonged incubation at $37^{\circ}$C.

As a side note, the lower specific activity values reported in Table~\ref{table:activity} compared with those of Table~\ref{table:stability} can be attributed to the prior incubation of the enzyme at pH 5 in citrate-phosphate buffer. The enzyme batch used for the measurement reported in Table~\ref{table:activity} had not been exposed to these conditions.

\begin{table}[t!]
    \centering
    \begin{tabular}{c c c c c}
    \hline
    Incubation time [h] && Specific activity [$\unit{U \, mg^{-1}}$]\\
    \hline
     $0.5$	&& $700 \pm 35$ \\ 
     $14$	&& $691 \pm 14$  \\ 
     $17$	&& $759 \pm 28$  \\ 
     $41$	&& $842 \pm 42$  \\ 
     $48$	&& $914 \pm 41$  \\
     $65$ && $757 \pm 50$  \\ 
     $72$	&& $739 \pm 36$  \\ 
      $161$	&& $606 \pm 13$  \\ 
    \hline
    \end{tabular}
    \caption{Specific activities of the labeled wild type enzyme, as determined using the DNS assay under reference conditions after different incubation times under the buffer and temperature conditions used in the Fluo-PCI experiments.}
    \label{table:stability}
\end{table}

\begin{figure}[H]
    \centering
    \includegraphics[width=0.8\columnwidth]{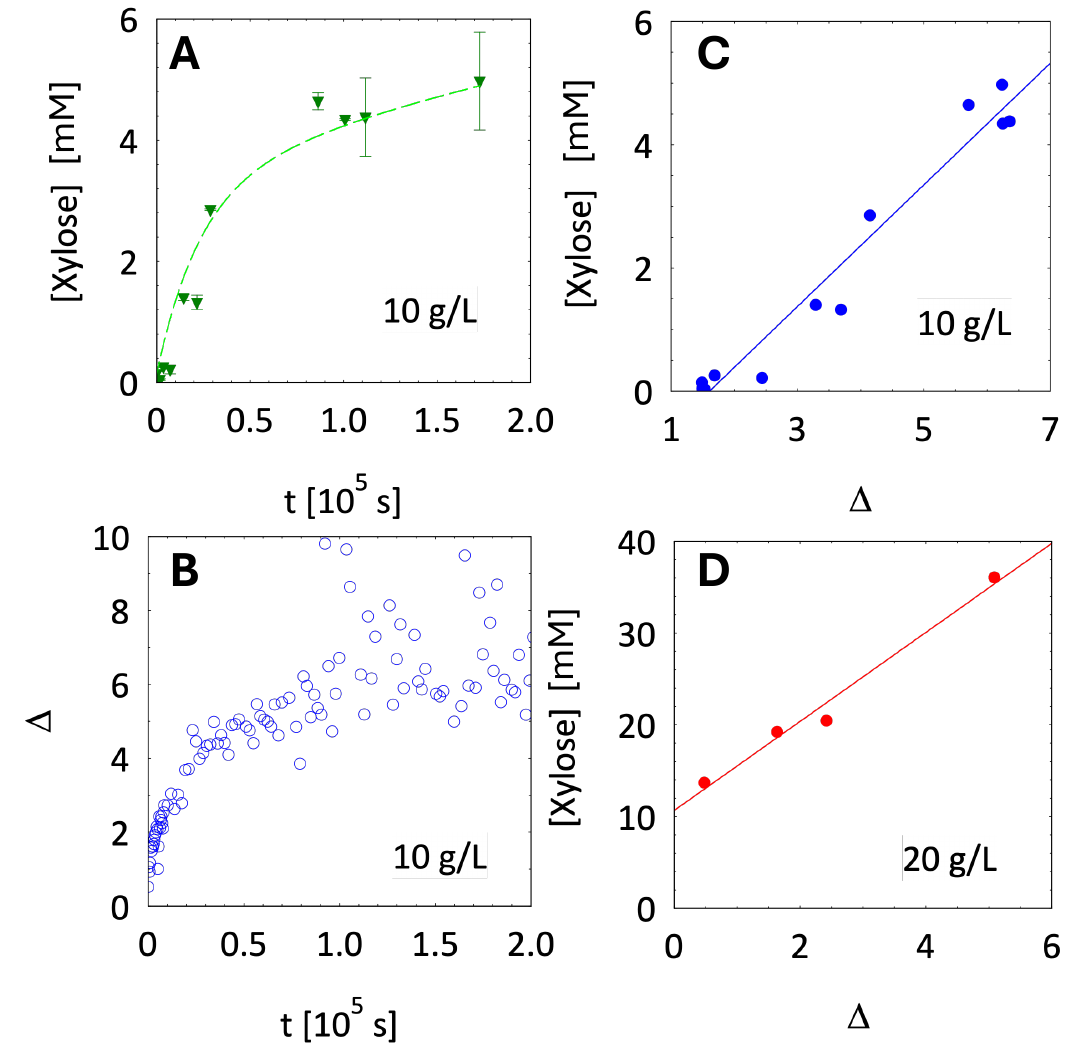}
    \caption{\textbf{Homogeneous degradation experiment with wild-type enzyme}. (A)~Xylose equivalent concentration \textit{vs} time measured by DNS assay (triangles) for a biopolymer FAX concentration $c = 10 \unit{g/L}$. The dashed line is a fit to the experimental data using the numerical solution to the MM equation with a decreasing fractional activity (Eq.~\ref{eq:S1}) and fixed parameters $k_{\alpha} = 4.6 \times 10^{-5} \unit{s^{-1}}$, $k_{\beta} = 2.7 \times 10^{-6} \unit{s^{-1}}$ and $\chi = 0.92$, obtained from the heterogeneous degradation data treatment (see main paper). Best fit parameters are $k_{cat} = 2.28 \unit{s^{-1}}$ and $K_m = 1.02 \times 10^{-3} \unit{mM}$. (B)~Biopolymer mobility $\Delta$ \textit{vs} time measured by time-resolved DLS at $T = 37 \celsius$ and $q = 22.2 \unit{\mu m^{-1}}$ for $c = 10 \unit{g/L}$. (C)~Biopolymer mobility \textit{vs}  Xylose equivalent concentration for the degradation experiment performed in homogeneous conditions at $c = 10 \unit{g/L}$ (data of Panels A and B, the enzyme concentration is $[E] = 65 \unit{nM}$). The blue line is a linear fit: [Xylose]$= A\Delta+B$, with $A = 0.987 \unit{mM}$ and $B = -1.584$. (D)~Biopolymer mobility \textit{vs}  Xylose equivalent concentration for the degradation experiment performed in homogeneous conditions with the second batch of enzyme (enzyme concentration $[E] = 38 \unit{nM}$, see text for details) at $c = 20 \unit{g/L}$ . The red line is a linear fit: [Xylose]$= A\Delta+B$, with $A = 4.853 \unit{mM}$ and $B = 10.660$. } 
    \label{fig:S7}
\end{figure}

\section*{Homogeneous degradation}

The long term kinetics of the degradation reaction of FAX biopolymer by fluorescently labeled WT enzymes has been quantified by DNS assay. Figure~\ref{fig:S7}A shows the result of the assay for a solution containing FAX chains (concentration $c = 10 \unit{g/L}$) and fluorescently labeled WT enzyme (concentration $E = 65 \unit{nM}$), during $48 \unit{hours}$. Note that, at the end of the experiment, the concentration of Xylose equivalents reaches $\sim 5 \unit{mM}$, a value somehow smaller than $\sim 40 \unit{mM}$, the theoretical number of possible cleavages if all X-X bonds of the FAX backbone were accessible to the enzyme active site. This lower experimental value can be explained by the fact that the enzyme can only hydrolyze xylosidic bonds linking two consecutive unsubstituted xylose residues~\cite{vardakou2008, rudjito2023}.

We characterized the homogeneous degradation of FAX in the same conditions also by time-resolved DLS performed at a scattering vector  $q = 22.2 \unit{\mu m^{-1}}$, from which we compute the time evolution of the polymer mobility $\Delta (t) = -\ln[{g_2(t; \Bar{\tau}=20 ms)-1}]$, shown in Fig.~\ref{fig:S7}B. The combination of the results at the same experimental times in Fig.~\ref{fig:S7}A and Fig.~\ref{fig:S7}B is shown in Fig.~\ref{fig:S7}D, where the concentration of Xylose equivalent is plotted as a function of the polymer microscopic mobility. The straight line is a fit to the data, demonstrating that the relationship between the concentration of degradation products and the mobility $\Delta$ is linear: [Xylose] =$ A\Delta +B$. Note that Fig.~\ref{fig:S7}C cannot be used as an absolute calibration for modeling the experiments performed at different enzyme concentrations, since these measurements are carried out at a FAX concentration of $c = 20 \unit{g/L}$, for which the coefficients $A$ and $B$ may, in principle, differ. A similar linear relation holds also for $c = 20 \unit{g/L}$ (see Fig.~\ref{fig:S7}D). However, data shown in Fig.~\ref{fig:S7}D have been obtained with a different batch of enzyme and cannot be used as an absolute calibration for modeling our experiments of directional degradation.


The dashed line in Fig.~\ref{fig:S7}A is a fit to the experimental points using the numerical solution to the MM equation with a decreasing fractional activity $f(t)$:
\begin{equation}
    \frac{\mathrm{d}[P]}{\mathrm{d}t}=k_{cat}[E] f(t)\frac{[S]_0 - [P]}{K_m+[S]_0 - [P]},
    \label{eq:S1}
\end{equation}
where $f(t)$ is given by Eq.~1 in the main text with the numerical values for  $k_{\alpha}$, $k_{\beta}$ and $\chi$, as derived from the fits of the directional degradation data, see main text. The good agreement with the data shown here, which were collected in a distinct experiment, with a different method and under spatially homogeneous conditions, further validate the modeling introduced in the main text. 

\section*{Histogram of the microscopic mobility kymograph}
\begin{figure}[h!]
    \centering
    \includegraphics[width=0.5\columnwidth]{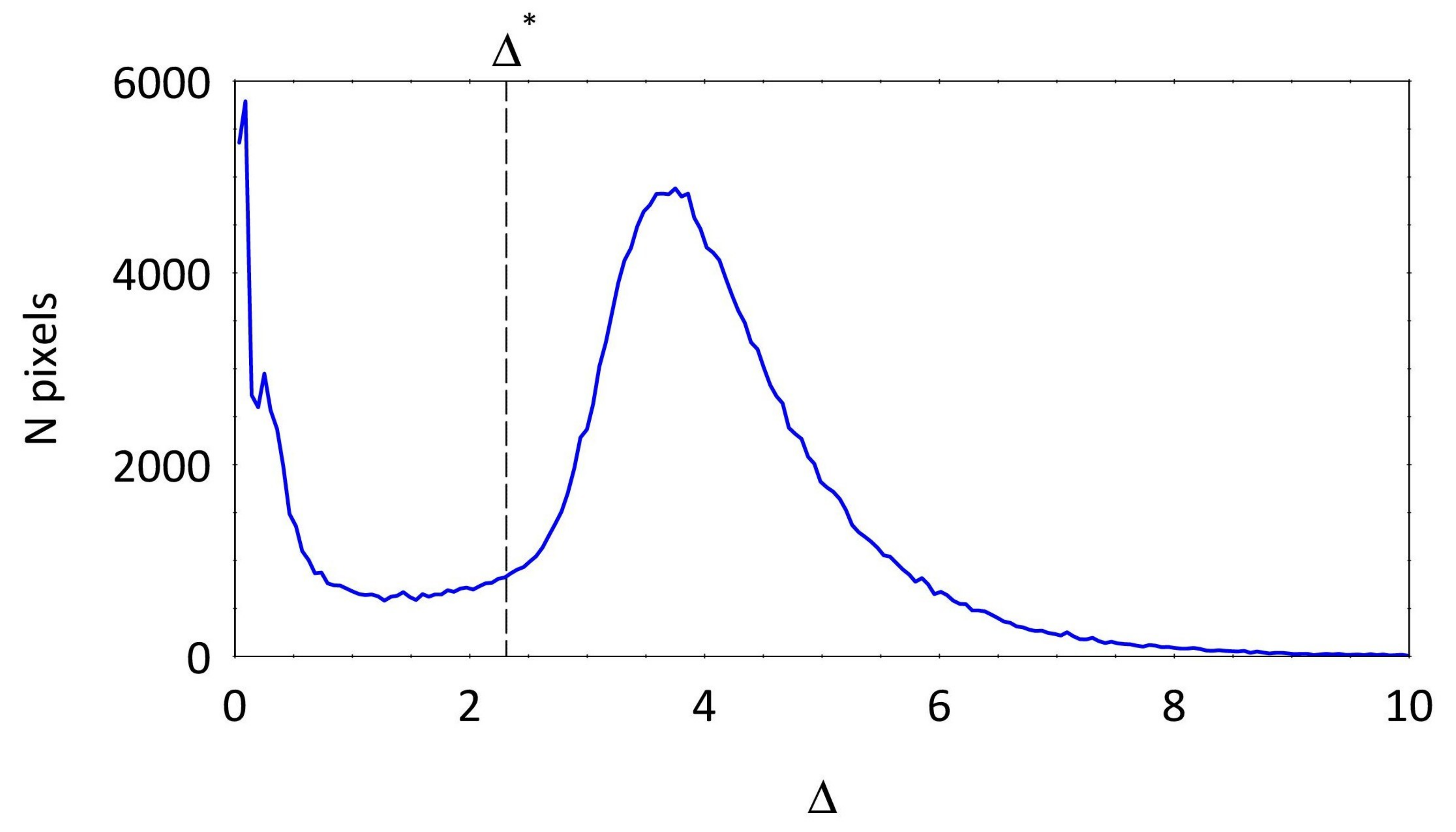}
    \caption{Histogram of the mobility kymograph, $\Delta (z,t)$, displayed in Fig.~2B in the main paper. The peaks at $\Delta \simeq 0$ and $\Delta \simeq 4$ correspond to low-mobility non-degraded biopolymer and high-mobility degraded biopolymer, respectively. The black dashed line indicates the threshold value $\Delta^*$ selected by the Otsu algorithm ($\Delta^* = 2.29$).}
    \label{fig:S1}
\end{figure}

The histogram of the biopolymer mobility kymograph shown in Fig.~2B of the main paper is displayed in Fig.~\ref{fig:S1}. The black dashed line indicates the threshold value $\Delta^* = 2.29$ selected by the Otsu algorithm, as described in the main paper.

\section*{Enzyme sizing and fits of enzyme diffusion}

\subsection*{Taylor Dispersion Analysis}

\begin{figure}[h!]
    \centering
    \includegraphics[width=0.5\columnwidth]{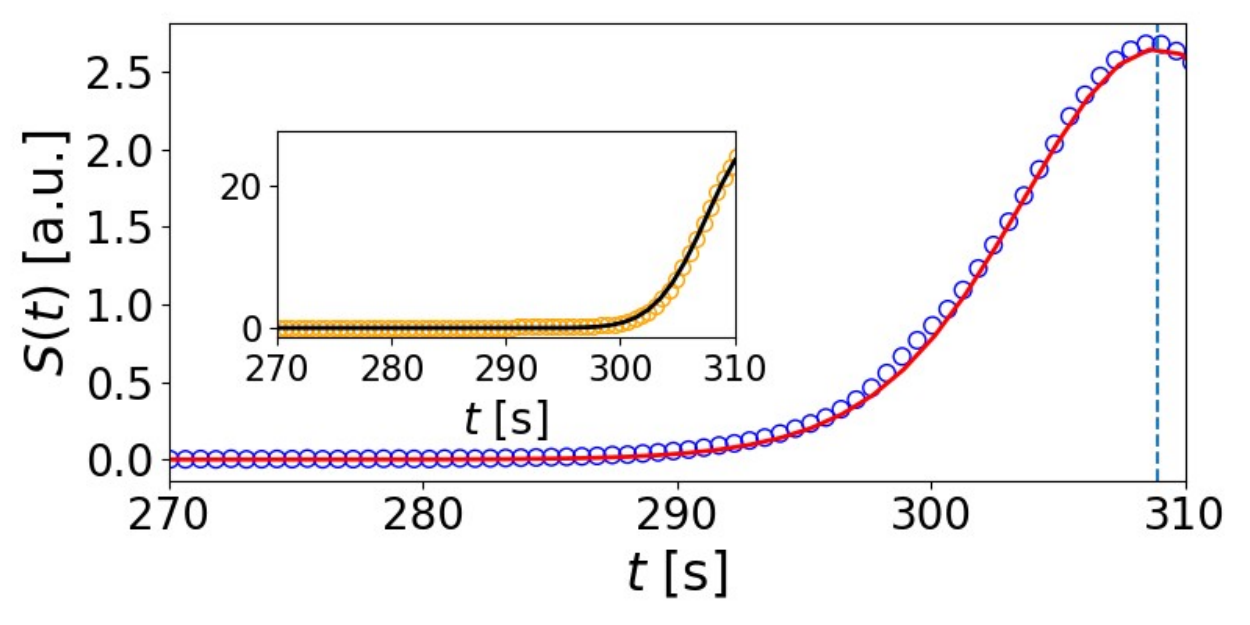}
    \caption{Taylorgram of the fluorescently labeled wild type enzyme solution (blue points). The red line is the best fit to the data using the two-species model (Eq.~\ref{eq:TDA2}), yielding $D_l = 1.1 \times 10^{-10} \unit{m^2/s}$, $D_s = 3.3 \times 10^{-10} \unit{m^2/s}$ and $r = 0.41$. Inset: Taylorgram of a solution of the free fluorophore (TRITC, orange points)  with a single-species fit (black line, Eq.~\ref{eq:TDA1}), yielding a diffusion coefficient $D = 4.1 \times 10^{-10} \unit{m^2/s}$.}
    \label{fig:S2}
\end{figure}

Taylor Dispersion Analysis (TDA) is a technique that measures the diffusion coefficients and therefore the hydrodynamic radii of particles and/or macromolecules in solution. TDA is based on the phenomenon of Taylor dispersion. Under suitable conditions~\cite{cottet2007, cipelletti2015}, when a suspension of monodisperse objects is injected in a capillary under laminar flow conditions, the combination of Brownian diffusion and of the Poiseuille radial flow profile leads to a distinctive temporal elution profile, called Taylorgram. The Taylorgram $S(t)$ measured, e.g. by absorbance, at a given position in the capillary may be described by the following equation:
\begin{equation}
    S(t) = \Sigma \sqrt{D} \exp\left[-12 D (t - t_0)^2/(R_c {t_0}^2) \right]
    \label{eq:TDA1}
\end{equation}
where $\Sigma$ is an instrumental constant, $R_c$ is the capillary radius, $D$ is the diffusion coefficient of the object in solution and $t_0$ is the peak time of the profile. In the case of a bidisperse suspension with large objects with diffusion coefficient $D_l$ and small objects with diffusion coefficient $D_s$ (the notation is the same as in the main paper), the elution profile reads:
\begin{equation}
    S(t) = \Sigma \left\{ r \sqrt{D_l} \exp\left[-k D_l (t - t_0)^2 \right] + (1 - r)\sqrt{D_s} \exp\left[-k D_s (t - t_0)^2 \right] \right\} 
    \label{eq:TDA2}
\end{equation}
with $k=12/(R_c {t_0}^2)$, and $r$ the (absorbance-weighted) relative abundance of the large species.

We exploit TDA to determine the hydrodynamic size of the fluorescent enzymes in solution. For details on the setup, we refer to~\cite{cottet2014}. The elution profile is measured by absorbance at wavelength $\lambda = 488 \unit{nm}$, thus exploiting the low-wavelength tail of the fluorescence absorption spectrum of the TRITC-labeled enzymes. The Taylorgrams are fitted considering both the one-species model (Eq.~\ref{eq:TDA1}) and the two-species model (Eq.~\ref{eq:TDA2}). To determine the best fit, we evaluate the Bayesian Information Criterion ($BIC$) parameter. The main graph of Fig.~\ref{fig:S2} shows the Taylorgram of an enzyme solution, along with the fit using the two species model (Eq.~\ref{eq:TDA2}). We find a $BIC$ for the two-species model ($BIC=-5240$) smaller than the one of the one-species model ($BIC=-2810$).  The extracted hydrodynamic radii are $R_l = 2.3 \unit{nm}$ and $R_s = 0.8 \unit{nm}$, with a relative number of large objects $r = 0.41$. Thus, TDA experiments show that two species are present in solution, whose size and relative concentration are comparable to those determined by fluorescence imaging of macroscopic diffusion in the biopolymer matrix (see main text). Furthermore, the Taylorgram measured for a sample with only free fluorophore (TRITC) in solution (inset of Fig.~\ref{fig:S2}) is best fitted with a single-species model ($BIC = -2810$ for the single-species model, and $BIC = -1890$ for the two-species model), yielding a hydrodynamic radius $R = 0.6 \unit{nm}$. This numerical value is comparable to $R_s$, the numerical value obtained for the small objects from TDA measurements of enzyme solution and from the diffusion experiments in the biopolymer matrix  with the Fluo-PCI setup. 

\subsection*{Models for the fluorescence signal}

\begin{figure}[t!]
    \centering
    \includegraphics[width=0.5\columnwidth]{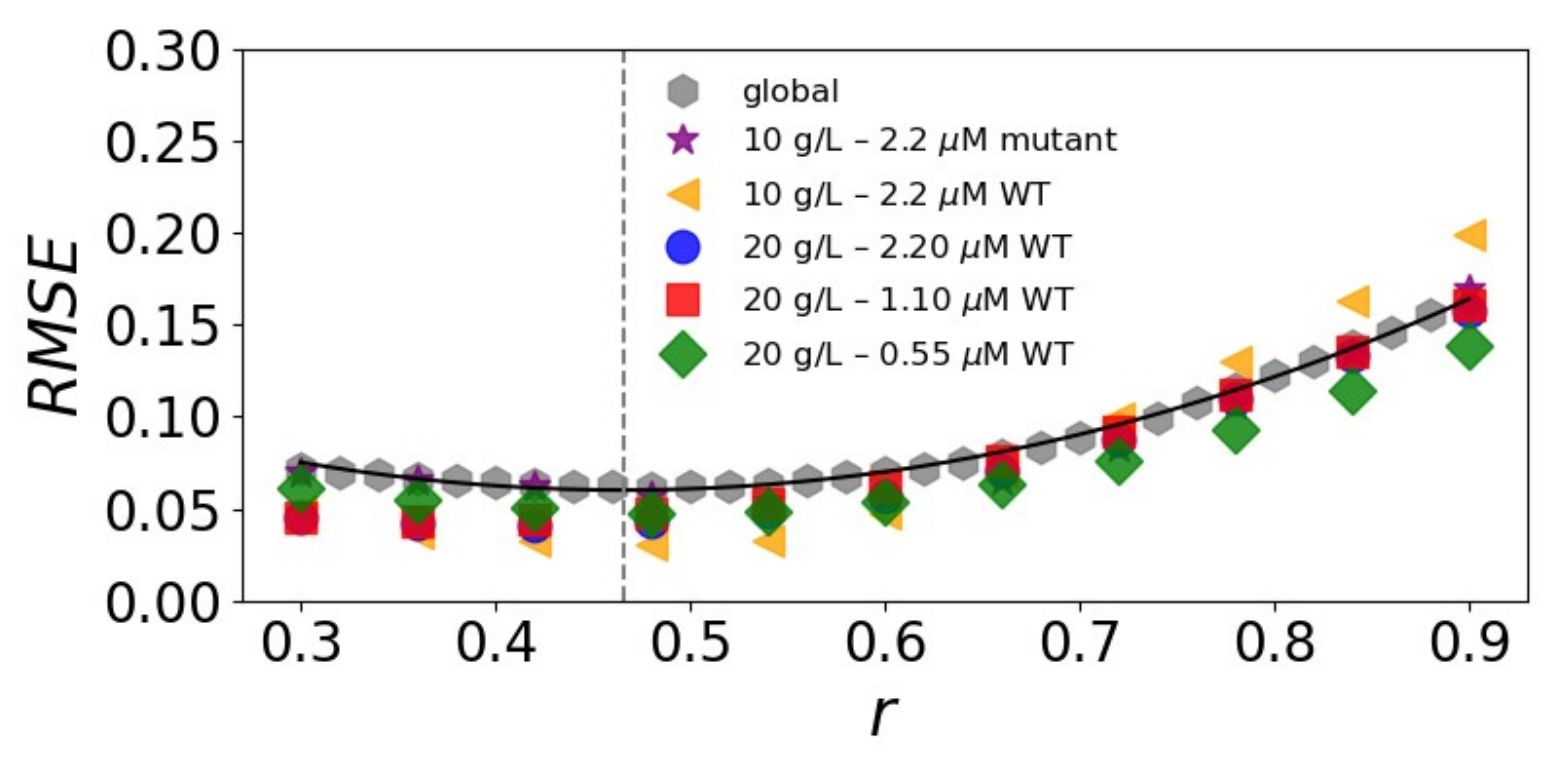}
    \caption{Root mean square error, $RMSE$, of the two-species diffusion fit adopted to model the fluorescence data collected by the fluorescence camera, as a function of $r$, the (intensity-weighted) relative number of large objects in solution. The global RMSE (grey hexagons simultaneous fit to all data) is minimized when $r = 0.47$, indicated by the dashed line. The black line is a parabolic fit to the data. For comparison, the $RMSE$ obtained from the fit to individual experimental datasets (see legend) is also shown.}
    \label{fig:S3}
\end{figure}

\begin{figure}[t!]
    \centering
    \includegraphics[width=\columnwidth]{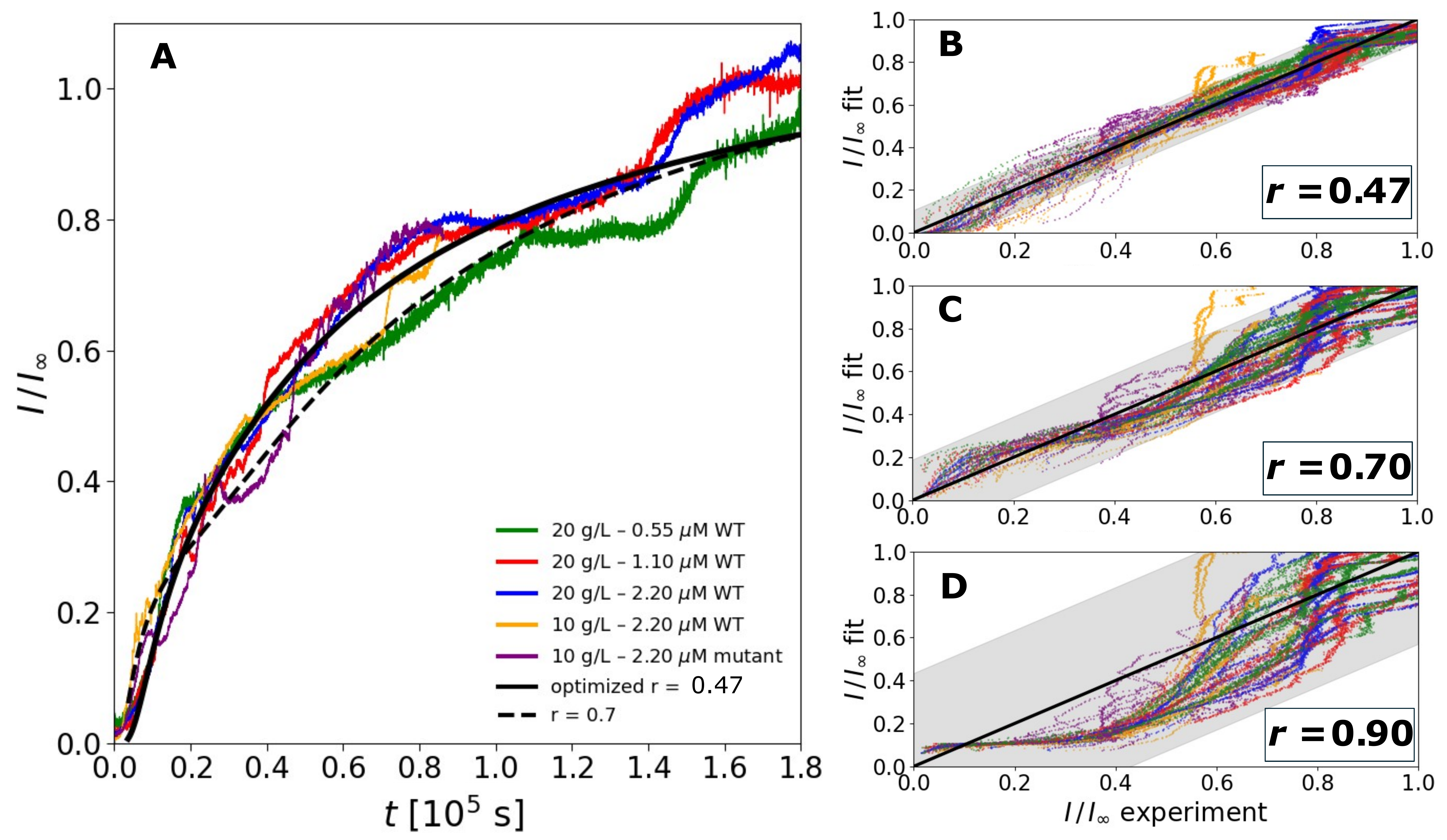}
    \caption{(A)~Time evolution of the normalized fluorescence intensity $I/I_{\infty}$ at $z= 6.22 \unit{mm}$ for different investigated samples (see legend). The black line is the best global fit according to Eqs.~MM4 and MM5 in the main text, with the optimized $r$ ($r=0.47$). The dash line corresponds to the best fit when imposing a test value, $r= 0.7$. (B, C, D)~Comparison between fit and experimental data at depths $z = 2.92, \, 3.68, \, 4.45, \, 4.95, \, 5.46, \, 6.22, \, 6.99, \, 7.49 \unit{mm}$, for $r = 0.47$ (B) (same as Fig.~4B in the main paper), $r= 0.70$ (C) and $r= 0.90$ (D). The deviation of the points from the black line of slope $1$ in (C) and (D) shows that the model is not able to fit satisfactorily the data sets as $r$ departs from its globally optimized value.}
    \label{fig:S4}
\end{figure}

Figure~\ref{fig:S3} shows the root mean square error, $RMSE$, of the fit to the data as a function of the (intensity-weighted) relative number of large objects in solution, $r$. For the simultaneous fit performed on data for all the investigated experimental conditions (grey hexagons), the minimum $RMSE$ is obtained for $r = 0.47 \pm 0.03$. This can be seen also from the residual plots in panels C and D of Fig.~\ref{fig:S4}, which display the deviation of the fit from the data for $r = 0.7$ and $r = 0.9$: the deviation clearly increases with $r$, showing the sensitivity of the fitting procedure. As an example, the global fit to the experimental data fixing $r = 0.7$ is showed in Fig.~\ref{fig:S4}A (dashed line).

\section*{Additional data for heterogeneous degradation}

\begin{figure}[h!]
    \centering
    \includegraphics[width=0.5\columnwidth]{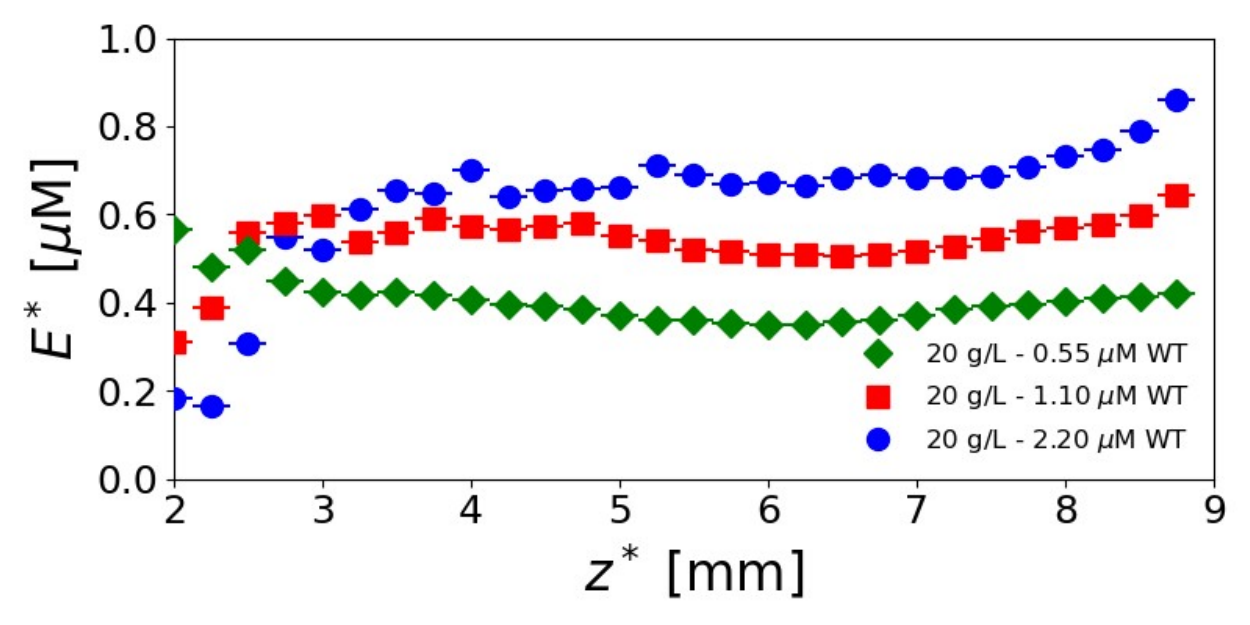}
    \caption{Enzyme concentration $E^*$ at the degradation front for experiments with biopolymer concentration $c = 20 \unit{g/L}$ and wild type enzyme with various final concentrations (see legend).}
    \label{fig:concentration_at_front}
\end{figure}

Figure~\ref{fig:concentration_at_front} displays the dependence of the enzyme concentration at the degradation front, $E^*$, on the front vertical position $z^*$, for three experiments with a same biopolymer concentration $c = 20 \unit{g/L}$ and various equilibrium final wild-type enzyme concentrations, $[E]_{\infty}$. Except at the top of the sample (small $z$, close to the initial biopolymer solution/enzyme solution interface), $E^*$ is roughly constant, regardless of the depth in the sample. However, $E^*$ increases as $[E]_{\infty}$ increases. This demonstrates that the degradation front is not defined by a threshold enzyme concentration.

\begin{figure}[h!]
    \centering
    \includegraphics[width=0.8\columnwidth]{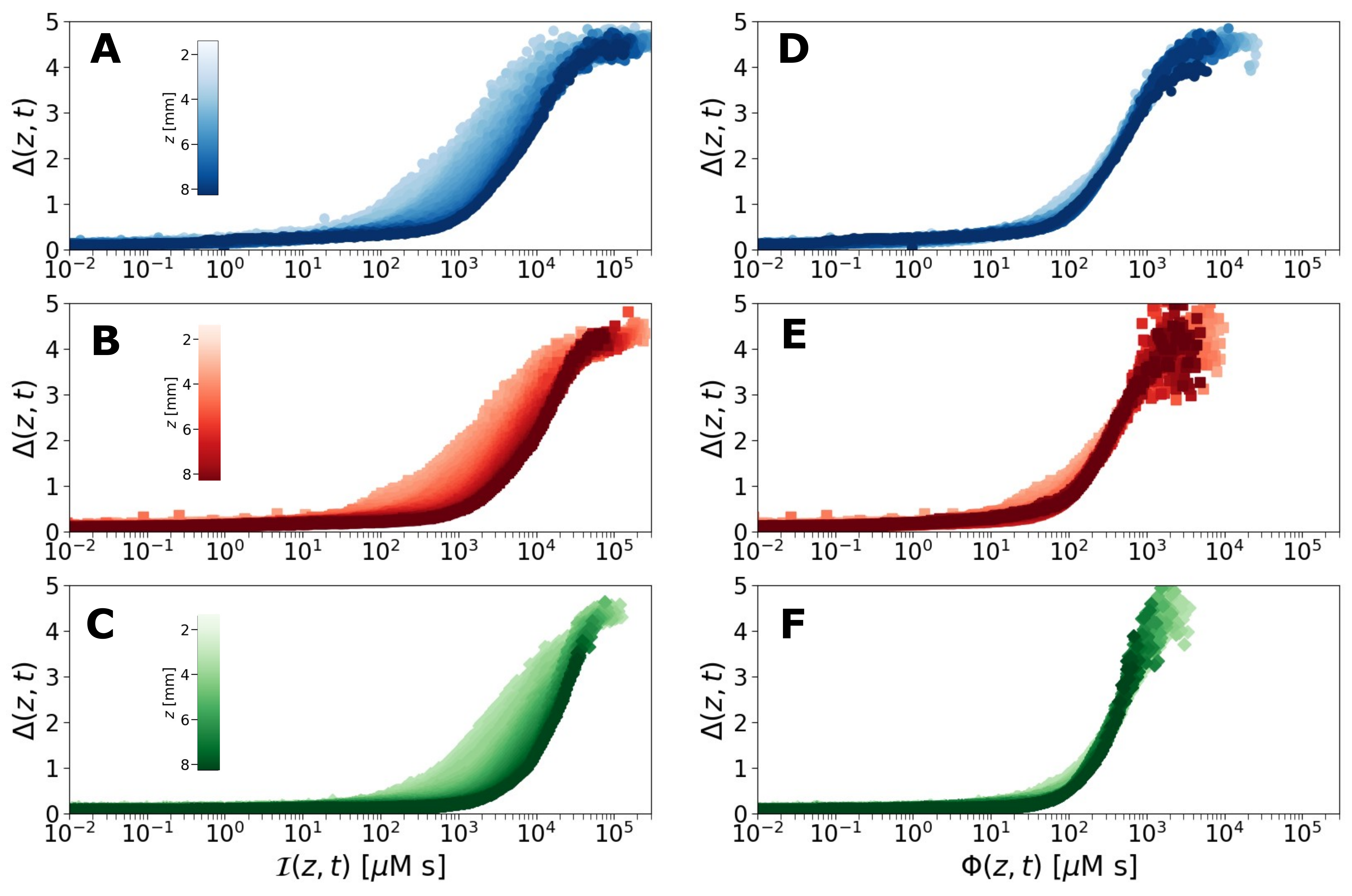}
    \caption{(A, B, C)~Mobility $\Delta$ as a function of $\mathcal{I} (z,t)$, the wild-type enzyme concentration at depth $z$ integrated  from time $0$ to time $t$ (see the main text for details), for samples with $c = 20 \unit{g/L}$ and (A)~$[E]_{\infty} = 2.2 \unit{\mu M}$; (B)~$[E]_{\infty} = 1.1 \unit{\mu M}$; (C)~$[E]_{\infty} = 0.55 \unit{\mu M}$. Each curve corresponds to a given $z$, from $z= 2.92$ to $z=8.00 \unit{mm}$ (see color bars). (D, E, F)~Mobility $\Delta$ as a function of $\Phi (z,t) = \int_0^t [E](z,t')f(t')\mathrm{d}t'$, accounting for enzyme deactivation (see the main text for details). Same samples and color code as in (A, B, C). Panels A and D are the same as Fig.~5A and Fig.~5D in the main paper, respectively. }
    \label{fig:S8}
\end{figure}

Figure~\ref{fig:S8} shows the biopolymer mobility $\Delta$ as a function of $\mathcal{I} (z,t)$, the integral of the wild-type enzyme concentration at depth $z$ from time $0$ to time $t$  (A,B,C), and as a function of $\Phi (z,t) = \int_0^t [E](z,t')f(t')\mathrm{d}t'$, for samples with $c = 20 \unit{g/L}$ and various enzyme concentrations.  $\Phi$ accounts for enzyme deactivation via the function $f (t) = \chi \exp{\left [-k_{\alpha}t \right ]}+(1-\chi)\exp{\left[-k_{\beta}t \right]}$ with $k_{\alpha} = 4.6 \times 10^{-5} \unit{s^{-1}}$, $k_{\beta} = 2.7 \times 10^{-6} \unit{s^{-1}}$ and $\chi = 0.92$, as obtained from a simultaneous fit of $\Delta(z,t)$ for the three experiments at $c = 20 \unit{g/L}$. Data measured at different $z$ do not overlap in (A,B,C) due to the enzyme deactivation, but do overlap when deactivation is considered (D,E,F), thus validating the model described in the main text.

\section*{Additional information for the modeling the heterogeneous enzymatic degradation of FAX}

\begin{figure}[h]
    \centering
    \includegraphics[width=0.5\columnwidth]{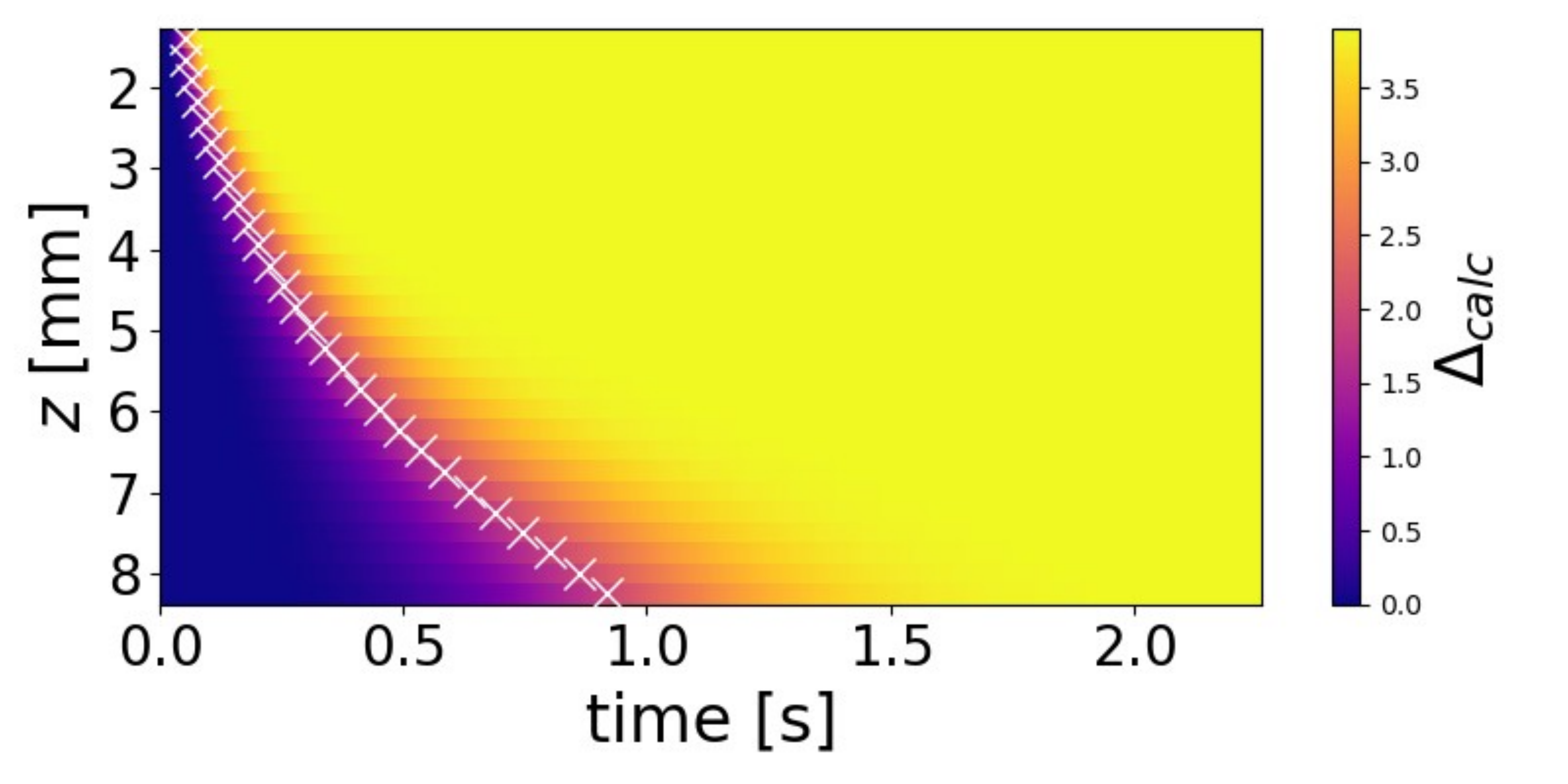}
    \caption{Calculated mobility kymograph generated using the solution of the time-dependent MM equation (Eq~.2 of the main text) for $c = 20 \unit{g/L}$ and $[E]_{\infty} = 2.2 \unit{\mu M}$ and the fitting parameters reported in the main text. The degradation front obtained using the Otsu algorithm is marked with white crosses.}
    \label{fig:S9}
\end{figure}

\begin{figure}[h]
    \centering
    \includegraphics[width=0.5\columnwidth]{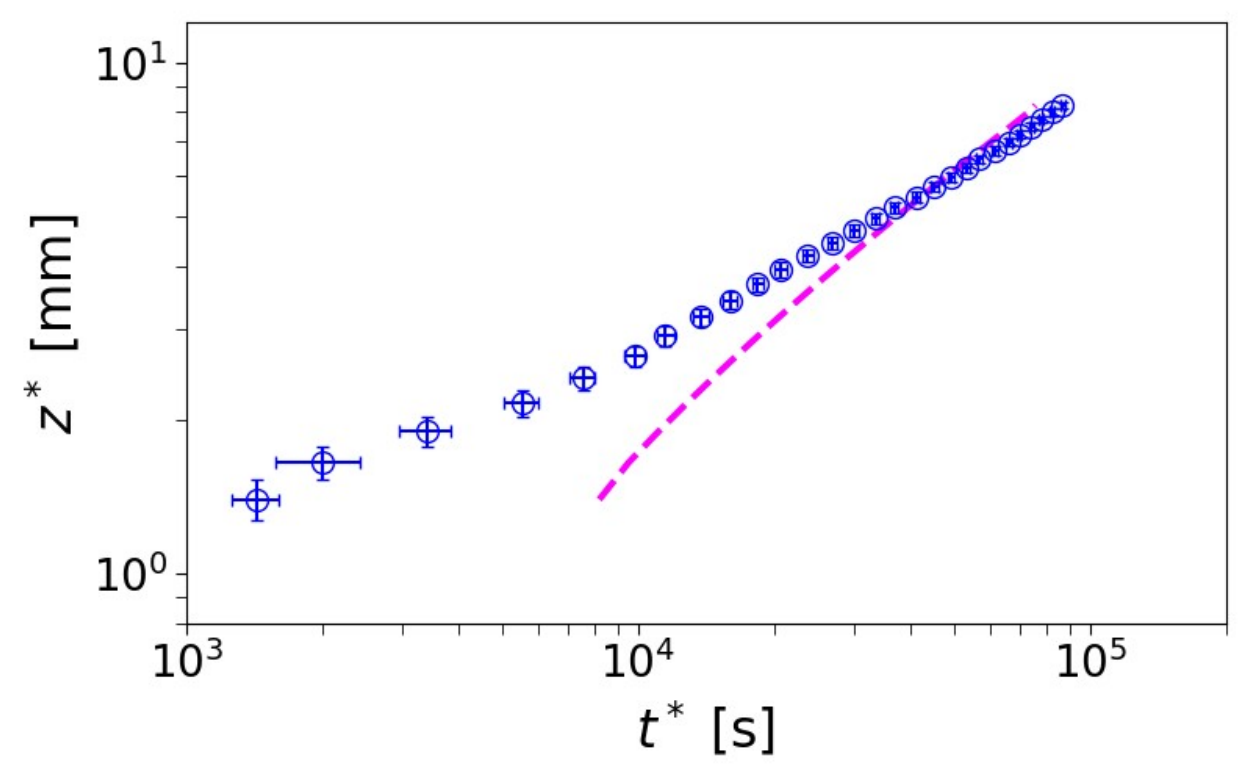}
    \caption{Log-log plot of the position of the degradation front as a function of time for biopolymer concentration $c = 20 g/L$ and WT enzyme concentration $[E]_\infty = 2.2 \unit{\mu M}$ (blue symbols). The pink dash line corresponds to a front obtained by fitting the kymographs with a model not accounting for enzyme deactivation ($f = 1$).}
    \label{fig:constant_activity}
\end{figure}

\begin{figure}[h]
    \centering
    \includegraphics[width=0.5\columnwidth]{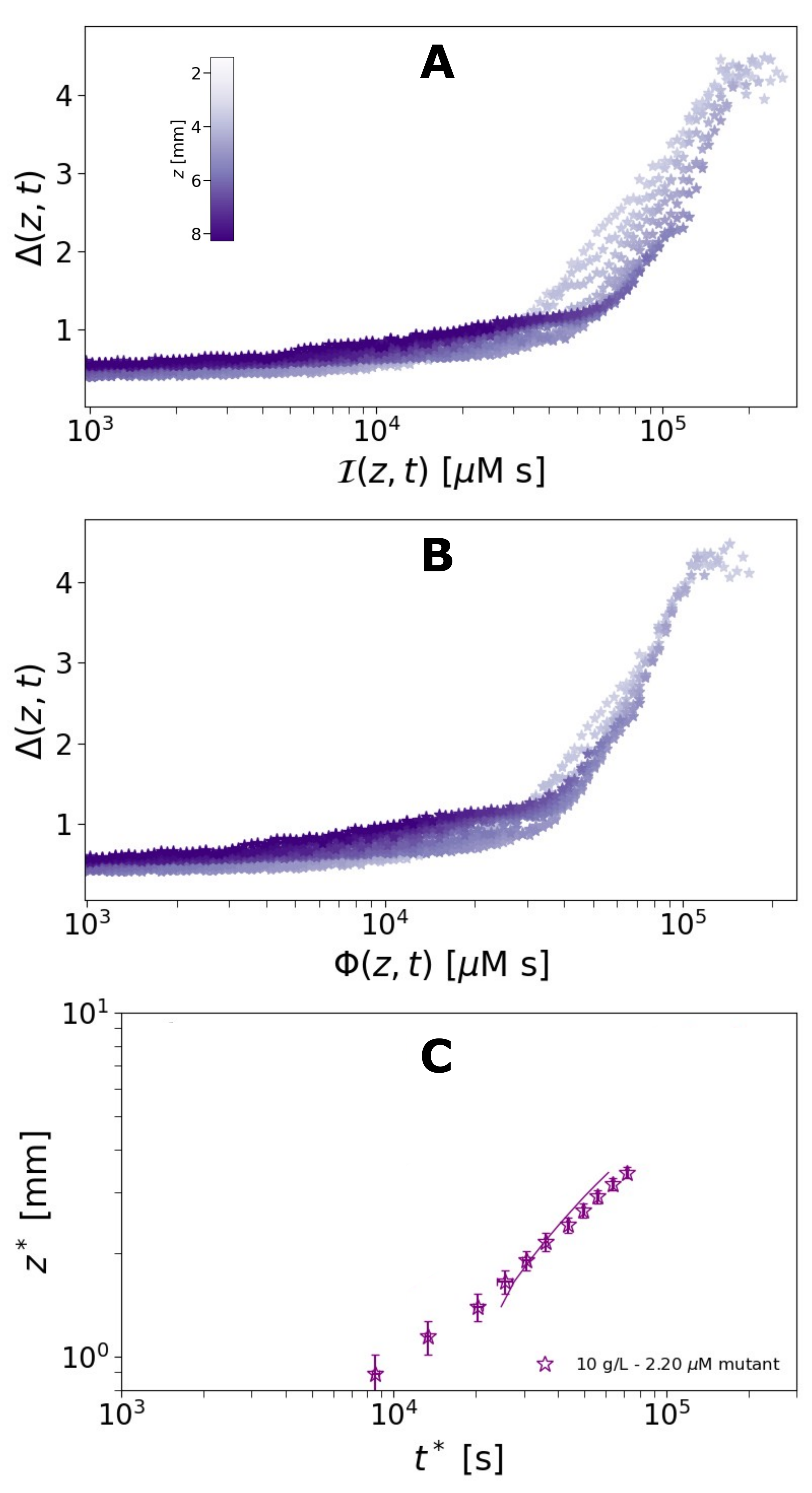}
    \caption{(A)~Mobility $\Delta$ as a function of $\mathcal{I}$ for samples with biopolymer concentration $c = 10 \unit{g/L}$ and  \textit{mutant} enzyme concentration $[E]_{\infty} = 2.2 \unit{\mu M}$. Each curve corresponds to a given $z$ (see color bar). (B)~$\Delta$ as a function of $\Phi (z,t) = \int_0^t [E](z,t')\exp{(-kt)}\mathrm{d}t'$ with the same color-code as in (A). For the mutant enzyme, the activity decrease is modelled by a simple exponential decay, with rate $k = 1.03 \times 10^{-5} \unit{s^{-1}}$. The overlap of the curves at different $z$s demonstrates the goodness of the model. (C) Degradation front obtained from the experimental data (symbols) and from the calculated mobility maps (line).}
    \label{fig:mutant_front}
\end{figure}

Figure~\ref{fig:S9} shows the computed mobility kymograph $\Delta_{calc} (z,t)$ for the sample with biopolymer concentration $c = 20 \unit{g/L}$ and wild-type final enzyme concentration $[E]_{\infty} = 2.2 \unit{\mu M}$, using the numerical solution to the MM-\textit{like} equation (Eq.~2 of the main paper), considering enzymes with diffusion coefficient $D_1 = 1.02 \times 10^{-10} \unit{m^2/s}$ and whose reduction of activity is modeled with a double-exponential (Eq.~1 of the main paper). The white crosses mark the degradation front times $t^*$ for each ROI reported as lines in Fig.~5C in the main paper.

Figure~\ref{fig:constant_activity} shows that the propagation of the degradation front is not satisfactorily modeled if enzyme deactivation is not considered.

Figure~\ref{fig:mutant_front}A shows the evolution of the biopolymer mobility $\Delta(z,t)$ as a function of $\mathcal{I}(z,t)$ for the mutant enzyme: data at different $z$ do not overlap, suggesting enzyme deactivation. As for the WT enzyme, we modeled deactivation by introducing an efficiency function $f$ and by replacing $\mathcal{I}$ by $\Phi$. In contrast to the wild type enzyme, here $f$ is satisfactorily described by a single exponential decay: $f(t) = \exp{(-kt)}$, with $k = 1.03 \times 10^{-5} \unit{s^{-1}}$, as shown in Fig.~\ref{fig:mutant_front}B in which $\Delta(z,t)$ is reported for different depths as a function of $\Phi(z,t)$. The degradation front retrieved from the calculated mobility maps is in good agreement with the experimental fronts, see Fig.~\ref{fig:mutant_front}C. However, the mutant enzyme exhibits a substantially slower decrease in activity during the degradation process, resulting in a behavior closer to the limiting case of a constant fractional activity. As a result, the corresponding exponent for the degradation front is closer to that obtained when no decay of enzymatic activity is considered, as shown in Fig.~\ref{fig:constant_activity}.



\end{document}